\documentclass{LMCS}

\newcommand{\lmon}{L^n_{k}}

\newcommand{\tup}[1]{\langle #1 \rangle}

\newcommand{\remove}[1]{}
\newcommand{\T}{{\mathcal{V}}}
\newcommand{\C}{{\mathcal{N}}}

\newcommand{\D}{{\mathcal{G}}}
\newcommand{\Q}{{\mathcal{Q}}}

\newcommand{\E}{{\mathcal{E}}}

\newcommand{\Oo}{{\mathcal{O}}}

\newcommand{\sig}{\sigma^*}

\newcommand{\Is}{I^*}

\newcommand{\Sss}{{\mathcal{S}}} 
\newcommand{\M}{{\mathcal{M}}}

\newcommand{\nin}{\not \in}
\newcommand{\term}{\textbf{t}}
\newcommand{\sterm}{\textbf{s}}

\newcommand{\vsg}{{v}}
\newcommand{\sm}{\sim^{S}}

\newlength{\defbaselineskip}
\newcommand{\setlinespacing}[1]%
           {\setlength{\baselineskip}{#1 \defbaselineskip}}
\newcommand{\doublespacing}{\setlength{\baselineskip}%
                           {1.3 \defbaselineskip}}

\usepackage{amsmath}
\usepackage{amsbsy}
\usepackage{amssymb}
\usepackage{proof,enumerate,hyperref}
\theoremstyle{plain}
\def\eg{{\em e.g.}}

\def\doi{4 (3:2) 2008}
\lmcsheading%
{\doi}
{1--23}
{}
{}
{Feb.~\phantom{0}1, 2007}
{Aug.~\phantom{0}6, 2008}
{}   
\begin{document}

\title[Canonical calculi with $(n,k)$-ary quantifiers]
{Canonical calculi with $(n,k)$-ary quantifiers}

\author[A.~Avron]{Arnon Avron} 
\address{Tel Aviv University}    
\email{\{aa,annaz\}@post.tau.ac.il}

\author[A.~Zamansky]{Anna Zamansky}   




\titlecomment{}


\begin{abstract}
  \noindent Propositional canonical Gentzen-type systems, introduced
  in 2001 by Avron and Lev, are systems which in addition to the
  standard axioms and structural rules have only logical rules in
  which exactly one occurrence of a connective is introduced and no
  other connective is mentioned.  A constructive coherence criterion
  for the non-triviality of such systems was defined and it was shown
  that a system of this kind admits cut-elimination iff it is
  coherent.  The semantics of such systems is provided using
  two-valued non-deterministic matrices (2Nmatrices).  In 2005
  Zamansky and Avron extended these results to systems with unary
  quantifiers of a very restricted form. In this paper we
  substantially extend the characterization of canonical systems to
  $(n,k)$-ary quantifiers, which bind $k$ distinct variables and
  connect $n$ formulas, and show that the coherence criterion remains
  constructive for such systems. Then we focus on the case of $k\in
  \{0,1\}$ and for a canonical calculus $G$ show that it is coherent
  precisely when it has a strongly characteristic 2Nmatrix, which in
  turn is equivalent to admitting strong cut-elimination.
\end{abstract}

\maketitle

\section*{Introduction}\label{S:one}

An {\em $(n,k)$-ary quantifier}\footnote{Generalized
quantifiers of this kind have been first considered in \cite{KM64}.
In \cite{Shroder} Natural Deduction calculi are provided
for $n$-place connectives and
quantifiers and it is shown that derivations
in such calculi are normalizable. } (for $n>0$, $k\geq 0$) is a generalized
logical connective, which binds $k$ variables and
connects $n$ formulas. Any $n$-ary propositional connective can be thought of as
an $(n,0)$-ary quantifier. For instance, the standard $\wedge$ connective binds no variables and
connects two
formulas: $\wedge(\psi_1,\psi_2)$. The standard first-order quantifiers
$\exists$ and
$\forall$ are $(1,1)$-quantifiers, as they bind one variable and connect one formula: $\forall x \psi,\exists x \psi$.
Bounded universal and existential quantifiers used in syllogistic
reasoning ($\forall x (p(x)\rightarrow q(x))$ and $\exists x
(p(x)\wedge q(x))$) can be represented as (2,1)-ary quantifiers $\overline{\forall}$ and $\overline{\exists}$,
binding one variable and connecting two formulas:
$\overline{\forall}x (p(x),q(x))$ and $\overline{\exists}x (p(x),q(x))$. An example of
$(n,k)$-ary quantifiers for $k>1$ are Henkin quantifiers\footnote{It should be noted that the semantic
interpretation of quantifiers
used in this paper is not sufficient for treating such quantifiers.} (\cite{Hen61,KM95}).
The simplest Henkin quantifier $Q_H$ binds 4 variables and connects one formula:
\[Q_H \ x_1 x_2 y_1 y_2\ \psi(x_1,x_2,y_1,y_2):= \begin{array}{cc}
  \forall x_1 & \exists y_1 \\
  \forall x_2 & \exists y_2
\end{array} \psi(x_1,x_2,y_1,y_2)\]
In this way of recording combinations of quantifiers, dependency
relations between variables are expressed as follows: an
existentially quantified variable depends on those universally
quantified variables which are on the left of it in the same row.

According to a long tradition in the philosophy of logic,
established by Gentzen in his classical paper {\em Investigations Into
Logical Deduction} (\cite{Gen69}), an ``ideal" set of introduction rules for a logical connective
should determine the {\em meaning} of the connective (see, \eg,
\cite{Zucker1,Zucker2}, and also \cite{Dosen} for a
general discussion). In \cite{AvrLev01,AvrLev04} the notion of
a ``canonical propositional Gentzen-type rule" was defined in precise
terms. A   constructive {\em coherence} criterion for
the non-triviality of systems consisting of such rules was provided,
and it was shown that a system of this
kind admits cut-elimination iff it is coherent. It was further proved
that the semantics of such systems is provided by
two-valued non-deterministic matrices (2Nmatrices), which form a
natural generalization of the classical matrix. In fact,
a characteristic 2Nmatrix was constructed for every coherent canonical
propositional system. 

In \cite{Zam_SL} the results were extended to systems (of a restricted
form) with unary quantifiers. A characterization of a ``canonical
unary quantificational rule" in such calculi was proposed (the
standard Gentzen-type rules for $\forall$ and $\exists$ are canonical
according to it), and a constructive extension of the coherence
criterion from \cite{AvrLev01,AvrLev04} for canonical systems of this
type was given. 2Nmatrices were extended to languages with unary
quantifiers, using a {\em distributional} interpretation of
quantifiers (\cite{Most,Carn87}). Then it was proved that again a
canonical Gentzen-type system of this type admits
cut-elimination\footnote{We note that by `cut-elimination' we mean
  here just the {\em existence} of proofs without (certain forms of)
  cuts, rather than an algorithm to transform a given proof to a
  cut-free one (for the assumptions-free case the term
  ``cut-admissibility" is sometimes used, but this notion is too weak
  for our purposes). See, \eg, \cite{BazL} for a resolution-based
  algorithm for cut-elimination in LK.}  iff it is coherent, and that
it is coherent iff it has a characteristic 2Nmatrix.

In this paper we make the intuitive notion of a ``well-behaved"
introduction rule for $(n,k)$-ary quantifiers formally precise. We
considerably extend the scope of the characterizations of
\cite{AvrLev01,AvrLev04,Zam_SL} to ``canonical $(n,k)$-ary
quantificational rules", so that both the propositional systems of
\cite{AvrLev01,AvrLev04} and the restricted quantificational systems
of \cite{Zam_SL} are specific instances of the proposed definition.
We show that the coherence criterion for
the defined systems remains decidable. Then we focus on the case
of $k\in \{0,1\}$ and show that the following statements
concerning a canonical calculus $G$ are equivalent: (i) $G$ is
coherent, (ii) $G$ has a strongly characteristic 2Nmatrix, and
(iii) $G$ admits strong cut-elimination. We show that coherence is
not a necessary condition for standard cut-elimination,
and then characterize a subclass of canonical systems for which
this property does hold.

\section{Preliminaries}\label{prelim}

\indent For any $n>0$ and $k\geq 0$, if a quantifier $\Q$ is of arity $(n,k)$, then
$\Q x_1...x_k
(\psi_1,...,\psi_n)$ is a formula whenever $x_1,...,x_k$ are
distinct variables and $\psi_1,...,\psi_n$ are formulas of $L$.

\noindent For interpretation of quantifiers, we use a generalized notion of
{\em distributions}
(see, e.g \cite{Most,Carn87}). Given a set $S$, $P^+(S)$ is the set of all
the nonempty subsets of $S$.
\begin{defi}\label{distri}Given a set of truth value $\T$, a {\em distribution}
of a (1,1)-ary quantifier $\Q$ is a function
$\lambda_\Q:P^+(\T)\rightarrow \T$.
\end{defi}
(1,1)-ary distribution quantifiers have been extensively studied
and axiomatized in many-valued logic. See, \eg, \cite{Carn87,Salzer96,Han98}.

In what follows, $L$ is a language with $(n,k)$-ary quantifiers,
that is with quantifiers $\Q_1,...,\Q_m$ with arities
($n_1,k_1$), ..., ($n_m,k_m$) respectively. Denote by
$Frm^{cl}_L$ the set of closed $L$-formulas and by $Trm^{cl}_L$
the set of closed $L$-terms. $Var=\{v_1,v_2,...,\}$ is the set of variables of
$L$. We use the metavariables $x,y,z$ to range over elements of
$Var$.

$\equiv_\alpha$ is the $\alpha$-equivalence relation between
formulas, i.e  identity up to the renaming of bound variables.
\begin{lem}\label{alphabkk} Let $\Q$ be an $(n,k)$-ary quantifier
of $L$ and $z_1,...,z_k$ fresh variables which do not occur in $\Q
x_1..x_k(\psi_1,...,\psi_n)$. Then:
$\Q x_1...x_k (\psi_1,...,\psi_n)\equiv_\alpha \Q
y_1...y_k(\psi'_1,...,\psi'_n)$ iff \\ $\psi_i\{z_1/x_1,...,z_k/x_k\}\equiv_\alpha
\psi'_i\{z_1/y_1,...,z_k/y_k\}$ for every $1\leq i \leq n$.

%

\end{lem}
\noindent The proof is not hard and is left to the reader.
\medskip

We use $[\ ]$ for application of functions in the
meta-language, leaving the use of $(\ )$ to the object language.
$A\{\term/x\}$ denotes the formula obtained from $A$ by
substituting $\term$ for $x$. Given an $L$-formula
$A$, $Fv[A]$ is the set of variables occurring free in
$A$. We denote $\Q x_1...x_k A$ by $\Q \overrightarrow{x} A$, and
$A(x_1,...,x_k)$ by $A(\overrightarrow{x})$.

A set of sequents $\Sss$ satisfies the {\em free-variable condition}
if the set of variables occurring bound in $\Sss$ is disjoint from the
set of variables occurring free in $\Sss$.

\section{Canonical Systems with (n,k)-ary quantifiers}
In this section we propose a precise characterization of a ``canonical
$(n,k)$-ary quantificational Gentzen-type rule".

Using an introduction rule for an $(n,k)$-ary quantifier $\Q$, we
should be able to derive a sequent of the form
$\Gamma\Rightarrow\Q x_1...x_k (\psi_1,...,\psi_n),\Delta$ or of the form
$\Gamma,Q x_1...x_k (\psi_1,...,\psi_n)\Rightarrow\Delta$, based
on some information about the subformulas of $\Q x_1...x_k
(\psi_1,...,\psi_n)$ contained in the premises of the rule. For
instance, consider the following standard rules for the (1,1)-ary
quantifier $\forall$:
\[\infer[(\forall\Rightarrow)]{\Gamma,\forall w\, A\Rightarrow
\Delta} {\Gamma,A\{\term/w\}\Rightarrow\Delta}\ \ \
\infer[(\Rightarrow \forall)]{\Gamma\Rightarrow\forall w\,
A,\Delta} {\Gamma\Rightarrow A\{z/w\},\Delta}\]where $\term,z$ are
free for $w$ in $A$ and $z $ does not occur free in the
conclusion. Our key observation is that the internal structure of
$A$, as well as the exact term $\term$ or variable $w$ used, are
immaterial for the meaning of $\forall$. What is important here is
the sequent on which $A$ appears, as well as whether
a term variable $\term$ or an eigenvariable $z$ is used. 

It follows that the internal structure of the formulas of $L$ used in
the description of a rule can be abstracted by using a simplified
first-order language, i.e., the formulas of $L$ in an introduction
rule of a $(n,k)$-ary quantifier, can be represented by {\em atomic}
formulas with predicate symbols of arity $k$. The case when the
substituted term is any $L$-term, will be signified by a constant, and
the case when it is a variable satisfying the above conditions - by a
variable. In other words, constants serve as term variables, while
variables are eigenvariables.

Thus in
addition to our original language $L$ with $(n,k)$-ary quantifiers we define another, simplified language.

\begin{defi}For $k\geq 0$, $n\geq 1$ and a set of constants $Con$,
$\lmon(Con)$ is the (first-order) language with $n$ $k$-ary predicate symbols
$p_1,...,p_n$ and the set of constants $Con$ (and no
quantifiers). The set of variables of $\lmon(Con)$ is
$Var=\{v_1,v_2,...,\}$.
\end{defi}

Note that $\lmon(Con)$ and $L$ share the same set of
variables. Furthermore, henceforth we assume that for every
$(n,k)$-ary quantifier $\Q$ of $L$, $\lmon(Con)$ is a subset of
$L$. This assumption is not necessary, but it makes the presentation
easier, as will be explained in the sequel.

Next we formalize the notion of a canonical rule and its
application.

\begin{defi}
Let $Con$ be some set of constants.
A {\em canonical quantificational rule of arity} $(n,k)$ is an
expression of the form $\{\Pi_i\Rightarrow\Sigma_i\}_{1\leq i \leq
m}/C$, where $m\geq 0$, $C$ is either $\Rightarrow
{\Q}v_1...v_k(p_1(v_1,...,v_k),...,p_n(v_1,...,v_k))$
or
${\Q}v_1...v_k(p_1(v_1,...,v_k),...,p_n(v_1,...,v_k))\Rightarrow$
for some $(n,k)$-ary quantifier ${\Q}$ of $L$ and for every $1\leq i
\leq m$: $\Pi_i\Rightarrow\Sigma_i$ is a clause\footnote{By a clause we mean a sequent containing only
atomic formulas.} over $\lmon(Con)$.
\end{defi}
Henceforth, in cases where the set of constants $Con$ is
clear from the context (it is the set of all constants occurring
in a canonical rule), we will write $\lmon$ instead of $\lmon(Con)$.

\noindent A canonical rule is a schematic representation, while for an actual application
we need to instantiate the schematic variables
by the terms and formulas of $L$. This is done using a mapping
function, defined as follows.
\begin{defi} \label{candefn}
Let $R=\Theta/C$ be an $(n,k)$-ary canonical rule, where $C$ is of one of the forms
$(\Q \overrightarrow{v} (p_1(\overrightarrow{v}),...,p_n(\overrightarrow{v}))\Rightarrow)$
or
$(\Rightarrow \Q \overrightarrow{v}
(p_1(\overrightarrow{v}),...,p_n(\overrightarrow{v})))$. Let $\Gamma$ be a set of
$L$-formulas and $z_1,...,z_k$ - distinct variables of $L$. An {\em
$\tup{R,\Gamma,z_1,...,z_k}$-mapping} is any function $\chi$ from
the predicate symbols, terms and formulas of $\lmon$ to formulas
and terms of $L$, satisfying the following conditions:
\begin{enumerate}[$\bullet$]
\item For every $1\leq i \leq n$, $\chi[p_i]$ is an $L$-formula.

\item $\chi[y]$ is a variable of $L$.
\item $\chi[x]\neq \chi[y]$ for every two variables $x\neq y$.
\item $\chi[c]$ is an $L$-term, such that $\chi[x]$ does not occur
in $\chi[c]$ for any variable $x$ occurring in $\Theta$.
\item For every $1\leq i \leq n$, whenever $p_i(\term_1,...,\term_k)$ occurs in $\Theta$, for
every $1\leq j \leq k$:
$\chi[\term_j]$ is a term free for $z_j$ in $\chi[p_i]$, and if
$\term_j$ is a variable, then $\chi[\term_j]$ does not occur free
in $\Gamma\cup\{\Q z_1...z_k (\chi[p_1],...,\chi[p_n])\}$.

\item $\chi[p_i(\term_1,...,\term_k)]
= \chi[p_i]\{\chi[\term_1]/z_1,...,\chi[\term_k]/z_k\}$.
\end{enumerate}
We extend $\chi$ to sets of
$\lmon(Con_\Theta)$-formulas as follows:
\[\chi[\Delta] = \{\chi[\psi]\ |\
\psi\in \Delta\}\]

\end{defi}
\medskip

Given a schematic representation of a rule and an instantiation
mapping, we can define an application of a rule as follows.

\begin{defi}

An {\em application} of a canonical rule of arity
$(n,k)$\hfill\break\medskip
\noindent$R=\{\Pi_i\Rightarrow\Sigma_i\}_{1\leq i \leq m}/{\Q}\overrightarrow{v}(
p_1(\overrightarrow{v}),...,p_n(\overrightarrow{v}))\Rightarrow$ is any
inference step of the form:
\[\infer[]{\Gamma,{\Q} z_1...z_k\ (\chi[p_1],...,\chi[p_n])
\Rightarrow \Delta} {\{\Gamma,\chi[\Pi_i]\Rightarrow
\Delta,\chi[\Sigma_i]\}_{1\leq i \leq m}}\] where $z_1,...,z_k$
are variables, $\Gamma\!,\Delta$ are any sets of $L$-formulas and
$\chi$ is some $\!\tup{R,\Gamma\cup\Delta,z_1,...,z_k}$- mapping.

An application of a canonical quantificational rule of the form
\[\{\Pi_i\Rightarrow\Sigma_i\}_{1\leq i \leq m}/\Rightarrow
{\Q}\overrightarrow{v}(
p_1(\overrightarrow{v}),...,p_n(\overrightarrow{v}))
\] 
is defined similarly.

\end{defi}
\noindent

\noindent Below we demonstrate the above definition by a number of
examples.
\begin{exas}\label{ex1}
\begin{enumerate}[(1)]
\item
The standard right introduction rule for $\wedge$, which can
be thought of as a $(2,0)$-ary quantifier is $\{\Rightarrow
p_1,\Rightarrow p_2 \}/\Rightarrow p_1\wedge p_2$. Its application
is of the form:
\[\infer[]{\Gamma\Rightarrow\psi_1\wedge \psi_2,\Delta}{\Gamma\Rightarrow \psi_1,\Delta & \Gamma\Rightarrow \psi_2,\Delta}\]
\item
  The standard introduction rules for the $(1,1)$-ary quantifiers
  $\forall$ and $\exists$ can be formulated as follows:\vskip-12 pt
\[\{p_1(c)\Rightarrow\}/\forall v_1\, p_1(v_1)\Rightarrow \ \ \ \
\{\Rightarrow p_1(v_1)\}/\Rightarrow\forall v_1\, p_1(v_1)\]
\[\{\Rightarrow p_1(d)\}/\Rightarrow \exists v_1\, p_1(v_1)\ \ \ \
\{p_1(v_1)\Rightarrow\}/\exists v_1\,p_1(v_1)\Rightarrow \]\vskip6 pt

\noindent Applications of these rules have the forms:
\[\infer[(\forall\Rightarrow)]{\Gamma,\forall w\, \psi\Rightarrow \Delta}
{\Gamma,\psi\{\term/w\}\Rightarrow\Delta} \ \ \ \infer[(\Rightarrow
\forall)]{\Gamma\Rightarrow\forall w\, \psi,\Delta}
{\Gamma\Rightarrow \psi\{z/w\},\Delta}\]\vskip-4 pt
\[\infer[(\Rightarrow\exists)]{\Gamma\Rightarrow \exists w\, A,\Delta}
{\Gamma \Rightarrow \psi\{\term/w\}, \Delta} \ \ \ \infer[(\exists\Rightarrow)]
{\Gamma,\exists w\, \psi\Rightarrow \Delta}
{\Gamma, \psi\{z/w\}\Rightarrow \Delta}\]
where $z$ is free for $w$ in
$\psi$, $z$ is not free in $\Gamma\cup \Delta\cup \{\forall w \psi\}$,
and $\term$ is any term free for $w$ in $\psi$.
\item
\label{ex3} Consider the bounded existential and universal
$(2,1)$-ary quantifiers $\overline{\forall}$ and
$\overline{\exists}$ (corresponding to $\forall x
.p_1(x)\rightarrow p_2(x)$ and $\exists x . p_1(x)\wedge p_2(x)$
used in syllogistic reasoning). Their corresponding rules can be
formulated as follows:
\[\{p_2(c)\Rightarrow\ ,\  \Rightarrow p_1(c)\}/\overline{\forall} v_1 \ (p_1(v_1),p_2(v_1))\Rightarrow\]
\[\{p_1(v_1)\Rightarrow
p_2(v_1)\}/\Rightarrow\overline{\forall} v_1\ (p_1(v_1),p_2(v_1))
\]
\[ \{p_1(v_1),p_2(v_1)\Rightarrow\}/\overline{\exists}\ v_1 (p_1(v_1),p_2(v_1))\Rightarrow\]\[\{\Rightarrow p_1(c)\ ,\ \Rightarrow
p_2(c)\}/\Rightarrow\overline{\exists} v_1 (p_1(v_1),p_2(v_1))\]

Applications of these rules are of the form:
\[\infer[]{\Gamma,\overline{\forall} z \ (\psi_1,\psi_2)\Rightarrow \Delta}
{\Gamma,\psi_2\{\term/z\}\Rightarrow \Delta & \Gamma\Rightarrow
\psi_1\{\term/z\},\Delta}\ \ \ \
\infer[]{\Gamma\Rightarrow\overline{\forall}z\
(\psi_1,\psi_2),\Delta}
{\Gamma,\psi_1\{y/z\}\Rightarrow\psi_2\{y/z\},\Delta}
\]\vskip-4 pt
\[\infer[]{\Gamma,\overline{\exists}z\ (\psi_1,\psi_2)\Rightarrow\Delta}
{\Gamma,\psi_1\{y/z\},\psi_2\{y/z\}\Rightarrow\Delta}\ \ \ \
\infer[] {\Gamma\Rightarrow\overline{\exists}z\
(\psi_1,\psi_2),\Delta }
{\Gamma\Rightarrow\psi_1\{\term/x\},\Delta &
\Gamma\Rightarrow\psi_2\{\term/x\},\Delta}\] where $\term$ and $y$
are free for $z$ in $\psi_1$ and $\psi_2$, $y$ does not occur free
in $\Gamma\cup\Delta\cup\{\overline{\exists} z (\psi_1,\psi_2)\}$.
\item \label{ex4}
Consider the (2,2)-ary rule \[\{p_1(v_1,v_2)\Rightarrow\ ,\  p_1(v_3,d)\Rightarrow
p_2(c,d)\}/\Rightarrow \Q v_1v_2(p_1(v_1,v_2),p_2(v_1,v_2))\] Its
application is of the form:
\[\infer[]{\Gamma\Rightarrow\Delta,\Q z_1z_2(\psi_1,\psi_2)}
{\Gamma,\psi_1\{w_1/z_1,w_2/z_2\}\Rightarrow \Delta &
\Gamma,\psi_1\{w_3/z_1,\term_1/z_2\}\Rightarrow\Delta,\psi_2\{\term_2/z_1,\term_1/z_2\}}\]
where $w_1,w_2,w_3,\term_1,\term_2$ satisfy the appropriate
conditions.
\end{enumerate}
\end{exas}

Note that although the derivability of the $\alpha$-axiom is
essential for any logical system, it is not guaranteed to be
derivable in a canonical system. What natural syntactic
conditions guarantee its derivability is still a question for
further research. For now we explicitly add the $\alpha$-axiom to the
canonical calculi.\medskip

\noindent{\bf Notation.} ({Following \cite{AvrLev01}, notations 3-5.}) Let $-t=f,
-f=t$ and $ite(t,A,B)=A$, $ite(f,A,B)=B$. Let $\Phi,A^s$ (where
$\Phi$ may be empty) denote $ite(s,\Phi\cup\{A\},\Phi)$. For
instance, the sequents $A\Rightarrow$ and $\Rightarrow A$ are
denoted by $A^{-s}\Rightarrow A^{s}$ for $s=f$ and $s=t$
respectively. According to this notation, a $(n,k)$-ary canonical rule is of the
form:\[\{\Sigma_j\Rightarrow \Pi_j\}_{1\leq j \leq m}/\Q
\overrightarrow{v}(p_1(\overrightarrow{v}),...,p_n(\overrightarrow{v}))^{-s}\Rightarrow \Q
\overrightarrow{v}(p_1(\overrightarrow{v}),...,p_n(\overrightarrow{v}))^{s}\] for $s\in
\{t,f\}$. For further abbreviation, we denote such rule by
$\{\Sigma_j\Rightarrow \Pi_j\}_{1\leq j \leq m}/\Q(s)$.

\begin{defi}\label{cano}
A Gentzen-type calculus $G$ is {\em canonical} if in addition to
the $\alpha$-axiom $A\Rightarrow A'$ for $A\equiv_\alpha A'$ and
the standard structural rules, $G$ has only canonical rules.
\end{defi}

\begin{defi}Two $(n,k)$-ary canonical introduction rules $\Theta_1/C_1$ and
$\Theta_2/C_2$ for $\Q$ are {\em dual} if for some $s\in \{t,f\}$:
 $C_1 = A^{-s}\Rightarrow A^s$ and $C_2 = A{^s}\Rightarrow
A^{-s}$, where $A = \Q v_1...v_k
(p_1(v_1,...,v_k),...,p_n(v_1,...,v_k))$.

\end{defi}
\remove{\begin{prop}\label{alpha}{\bf ($\alpha$-conversion)} Let
$\Gamma'\Rightarrow\Delta'$ be an $\alpha$-variant of
$\Gamma\Rightarrow\Delta$. Then any derivation  of
$\Gamma\Rightarrow\Delta$ in a canonical calculus $G$ can be
converted into a derivation in $G$ of $\Gamma'\Rightarrow\Delta'$
by just renaming of bound variables.

\end{prop}

\begin{prop}\label{substlemma} {\bf (Substitution)} Let
$G$ be a canonical calculus. Let \\ $\Gamma\Rightarrow\Delta$ be a
sequent derivable in $G$. For any $\term$ free for $x$ in any
$\psi\in\Gamma\cup\Delta$,
$\Gamma\{\term/x\}\Rightarrow\Delta\{\term/x\}$ is derivable in
$G$ with the same derivation height.

\end{prop}
The proofs of the above propositions are rather standard, by
induction on the length of the proofs.\\}

Although we can define arbitrary canonical systems using our
simplified language $\lmon$, our quest is for systems, the syntactic
rules of which define the semantic meaning of logical
connectives/quantifiers. Thus we are interested in calculi with a
``reasonable" or ``non-contradictory" set of rules, which allows for
defining a sound and complete semantics for the system.  This can be
captured syntactically by the following extension of the {\em
  coherence} criterion of \cite{AvrLev01,Zam_SL}.

\begin{defi}\label{renaming}For two sets of clauses $\Theta_1,\Theta_2$ over $\lmon$, ${\sf
Rnm}(\Theta_1\cup\Theta_2)$ is a set $\Theta_1\cup\Theta'_2$,
where $\Theta'_2$ is obtained from $\Theta_2$ by a fresh renaming of
constants and variables which occur in $\Theta_1$.

\end{defi}

Henceforth it will be convenient (but not essential) to assume that the fresh
constants used for the renaming are in $L$.

\begin{defi}\label{cohdefn} {\bf (Coherence)}\footnote{
A strongly related coherence criterion is defined in \cite{Miller}, where
linear logic is used to reason about various sequent systems. Our
coherence criterion is also equivalent in the context of canonical calculi to the reductivity
condition in \cite{Ciabnk}, as will be explained in the sequel.}
A canonical calculus $G$ is {\em coherent} if for every two
dual canonical rules $\Theta_1/\Rightarrow A$ and $\Theta_2/A\Rightarrow$,
the set of clauses ${\sf Rnm}(\Theta_1\cup
\Theta_2)$ is classically inconsistent.
\end{defi}

Note that the principle of renaming of clashing constants and variables is
similar to the one used in first-order resolution. The importance
of this principle for the definition of coherence will be
explained in the sequel.

\begin{prop}\label{decidability} {\bf (Decidability of coherence)} The coherence of a canonical calculus
$G$ is decidable.
\end{prop}
\proof The question of classical consistency of a finite set
of clauses without function symbols (over $\lmon$) can be shown to be
equivalent to satisfiability of a finite set of universal formulas
with no function symbols. This is decidable (by an obvious application of Herbrand's theorem).

%
\section{The semantic framework}

\subsection{Non-deterministic matrices} Our main
semantic tool are non-deterministic matrices (Nmatrices),
first introduced in \cite{AvrLev01,AvrLev04} and extended
in \cite{Zam_ISMVL,Zam_SL}. These structures are a generalization of the
standard concept of a many-valued matrix, in which the
truth-value of a formula is chosen non-deterministically from a
given non-empty set of truth-values. Thus, given a set of
truth-values $\T$, we can generalize the notion of a distribution
function of an $(n,k)$-ary quantifier $\Q$ (from Definition. \ref{distri})
to a function $\lambda_\Q:P^+(\T^n)\rightarrow P^+(\T)$. In other
words, given some distribution $Y$ of n-ary vectors of truth values, the interpretation function non-deterministically chooses the
truth value assigned to $\Q \overrightarrow{z}(\psi_1,...,\psi_n)$
out from $\lambda_\Q[Y]$ .

\begin{defi} {\bf (Non-deterministic matrix)}
A non-deterministic matrix \\(henceforth
Nmatrix) for $L$ is a tuple $\M = <\T,\D,\Oo>$, where:
\begin{enumerate}[$\bullet$]
\item $\T$ is a non-empty set of truth values.
\item $\D$ (designated truth values) is a non-empty proper subset of $\T$.
\item $\Oo$ is a set of interpretation functions: for every $(n,k)$-ary quantifier $\Q$ of L,
$\Oo$ includes the corresponding distribution function
$\tilde{\Q}_\M:P^+({\T}^n)\rightarrow P^+(\T)$.

\end{enumerate}
\end{defi}
Note the special treatment of propositional
connectives in the definition above. In \cite{AvrLev01,Zam_SL},
an Nmatrix includes an interpretation function
$\tilde{\diamond}:\T^n\rightarrow P^+(\T)$ for every $n$-ary
connective of the language; given a valuation $v$, the truth value
$v[\diamond(\psi_1,...,\psi_n)]$ is chosen non-deterministically
from $\tilde{{\diamond}}[\tup{v[\psi_1],...,v[\psi_n]}]$. In the
definition above, the interpretation of a propositional connective
$\diamond$ is a function of another type:
$\tilde{\diamond}:P^+(\T^n)\rightarrow P^+(\T)$. This can be
thought as a generalization of the previous definition,
identifying the tuple $\tup{v[\psi_1],...,v[\psi_n]}$ with the
singleton $\{\tup{v[\psi_1],...,v[\psi_n]}\}$. The advantage of
this generalization is that it allows for a uniform treatment of
both quantifiers and propositional connectives.

\begin{defi}{\bf (L-structure)} Let $\M$ be an Nmatrix for $L$. An
L-structure for $\M$ is a pair $S=\tup{D,I}$ where $D$ is a
(non-empty) domain and $I$ is a function interpreting constants,
predicate symbols and function symbols of $L$, satisfying the
following conditions: $I[c]\in D$, $I[p^n]: D^n\rightarrow \T$ is
an n-ary predicate, and $I[f^n]: D^n\rightarrow D$ is an n-ary
function. \\$I$ is extended to interpret closed terms of $L$ as
follows:
$$I[f(\term_1,...,\term_n)] = I[f][I[\term_1],...,I[\term_n]]$$
\end{defi}

Here a note on our treatment of quantification in the framework of
Nmatrices is in order. The standard approach to interpreting quantified formulas
is by using {\em objectual} (or referential) semantics, where
the variable
is thought of as ranging
over a set of objects from the domain (see, \eg,
\cite{End72,Fit96}). An alternative approach
is {\em substitutional} quantification (\cite{Leb01}), where
quantifiers are interpreted substitutionally, i.e. a universal (an existential)
quantification is true if and only if every one (at least one) of its substitution
instances is true (see, \eg, \cite{Sho67,Dal97}).
\cite{Zam_ISMVL} explains the motivation behind choosing the substitutional
approach for the framework of Nmatrices, and points out the problems
of the objectual approach in this context. The substitutional
approach assumes that every element of the domain has a closed term
referring to it. Thus given a structure $S=\tup{D,I}$, we extend the language $L$ with {\em individual
constants}, one for each element of $D$.

\begin{defi}\label{ld} {\bf ( L(D) ) }Let S=$\tup{D,I}$ be an $L$-structure for an
Nmatrix $\M$. $L(D)$ is the language obtained from $L$ by adding
to it the set of {\em individual constants} $\{\overline{a}\ |\
a\in D\}$. $S' = \tup{D,I'}$ is the $L(D)$-structure, such that
$I'$ is an extension of $I$ satisfying: $I'[\overline{a}]=a$.
\end{defi}

Given an $L$-structure $S=\tup{D,I}$, we shall refer to
the extended $L(D)$-structure $\tup{D,I'}$ as $S$ and to $I'$ as
$I$ when the meaning is clear from the context.
\remove{\begin{lem}\label{const}
\begin{enumerate}[(1)]
\item If $\psi\equiv_\alpha\psi'$, then for any constant $c$ of $L(D)$
and any variable $x$, $\psi\{c/x\}\equiv_\alpha\psi'\{c/x\}$.
\item If $\psi\{c'/x\}\equiv_\alpha\psi'\{c'/y\}$ for some
constant $c'$ of $L(D)$ not occurring in $\psi,\psi'$, then $Q x
\psi\equiv_\alpha Q y \psi'$.
\item If $\psi\equiv_\alpha\psi'$ and
$\varphi\equiv_\alpha\varphi'$, then
$\psi\diamond\varphi\equiv_\alpha\psi'\diamond\varphi'$.

\end{enumerate}
\end{lem}
}

\begin{defi}{\bf ($S$-substitution)}
Given an $L$-structure $S=\tup{D,I}$ for an Nmatrix $\M$ for $L$,
an {\em $S$-substitution} is a function $\sigma:Var\rightarrow
Trm^{cl}_{L(D)}$. It is extended to $\sigma:Trm_L\cup Frm_L\rightarrow
Trm^{cl}_{L(D)}\cup Frm^{cl}_{L(D)}$ as follows: for a term $\term$ of
$L(D)$, $\sigma[\term]$ is the closed term obtained from $\term$
by replacing every $x\in Fv[\term]$ by $\sigma[x]$. For a formula
$\varphi$, $\sigma[\varphi]$ is the sentence obtained from
$\varphi$ by replacing every $x\in Fv[\varphi]$ by
$\sigma[x]$.\\
Given a set $\Gamma$ of formulas, we denote the set
$\{\sigma[\psi]\ |\ \psi\in \Gamma\}$ by
$\sigma[\Gamma]$.\end{defi}
\remove{\begin{lem}\label{sigal} Let
$S$ be an $L$-structure and $\sigma$ an $S$-substitution. For
every two $L$-formulas $\psi,\psi'$: if $\psi\equiv_\alpha \psi'$,
then $\sigma[\psi]\equiv_\alpha \sigma[\psi']$.

\end{lem}
}

The motivation for
the following definition is purely technical and is related to extending the language
with the set of individual constants $\{\overline{a}\ |\ a\in
D\}$. Suppose we have a closed term $\term$, such that $I[\term]=a\in
D$. But $a$ also has an individual constant $\overline{a}$ referring to
it. We would like to be able to substitute $\term$ for
$\overline{a}$ in every context.

\begin{defi}{\bf (Congruence
of terms and formulas)} Let $S$ be an $L$-structure for an Nmatrix
$\M$. The relation $\sim^S$ between terms of $L(D)$ is defined
inductively as follows:
\begin{enumerate}[$\bullet$]
\item $x\sim^S x$
\item For closed terms $\term,\term'$ of $L(D)$:
$\term\sim^S \term'$ when $I[\term]=I[\term']$.
\item If $\term_1\sim^S\term'_1,...,\term_n\sim^S\term'_n$, then
$f(\term_1,...,\term_n)\sim^S f(\term'_1,...,\term'_n)$.
\end{enumerate}
The relation $\sm$ between formulas of $L(D)$ is defined as
follows:
\begin{enumerate}[$\bullet$]
\item If $\term_1\sim^S\term'_1,\term_2\sim^S\term'_2,...,
\term_n\sim^S\term'_n$, then
$p(\term_1,...,\term_n)\sm p(\term'_1,...,\term'_n)$.
\item If $\psi_1\{\overrightarrow{z}/\overrightarrow{x}\}\sm\varphi_1\{\overrightarrow{z}
/\overrightarrow{y}\}, ...,
\psi_n\{\overrightarrow{z}/\overrightarrow{x}\}\sm
\varphi_n\{\overrightarrow{z}/\overrightarrow{y}\}$, where
$\overrightarrow{x}=x_1...x_k$ and $\overrightarrow{y}=y_1...y_k$
are distinct variables and $\overrightarrow{z}=z_1...z_k$ are new
distinct variables, then $\Q \overrightarrow{x}
(\psi_1,...,\psi_n)\sm \Q \overrightarrow{y}
(\varphi_1,...,\varphi_n)$ for any $(n,k)$-ary quantifier $\Q$ of
$L$.
\end{enumerate}

\end{defi}

Intuitively, $\psi\sm\psi'$ if $\psi'$ can be obtained from $\psi$ by
possibly renaming bound variables and by any number of substitutions
of a closed term $\term$ for another closed term $\sterm$, so that
$I[\term]=I[\sterm]$. 
$\equiv_\alpha\subseteq \sm$.

\remove{
\noindent The following lemma shows that the $\sm$ relation is
independent of the choice of the new variable $z$.

\begin{lem}\label{mainres} Let $S$ be an L-structure for an Nmatrix $\M$. Let
$\term,\term'$ be terms and $\psi,\psi'$ formulas of $L(D)$. Let
$z$ be a new variable and $\sterm$ - a term of $L(D)$ which is (i)
free for $x$ in $\term$ and $\psi$, and (ii) free for $y$ in
$\term'$ and $\psi'$.
\begin{enumerate}[$\bullet$]
\item If $\term\{z/x\}\sim^S\term'\{z/y\}$,
then $\term\{\sterm/x\}\sim^S\term'\{\sterm/y\}$.

\item If $\psi\{z/x\}\sm\psi'\{z/y\}$,
then \\$\psi\{\sterm/x\}\sm\psi'\{\sterm /y\}$.

\end{enumerate}
\end{lem}
}

\begin{lem}(\cite{Zam_ISMVL})\label{solves}
Let $S$ be an $L$-structure for an Nmatrix $\M$. Let $\psi,\psi'$
be formulas of $L(D)$. Let $\term,\term'$ be closed terms of
$L(D)$, such that $\term\sm\term'$.
\begin{enumerate}[\em(1)]
\item If $\psi\equiv_\alpha\psi'$, then $\psi\sm\psi'$.
\item If $\psi\sm\psi'$, then $\psi\{\term/x\}\sm \psi'\{\term'/x\}$.
\end{enumerate}

\end{lem}

\begin{defi}{\bf (Legal valuation)}
Let $S=\tup{D,I}$ be an $L$-structure for an Nmatrix $\M$. An
$S$-valuation $v:Frm^{cl}_{L(D)}\rightarrow\T$ is {\em legal in $\M$}
if it satisfies the following conditions:
\begin{enumerate}[(1)]
\item $v[\psi] = v[\psi']$ for every two
sentences $\psi,\psi'$ of $L(D)$, such that $\psi\sm\psi'$.
\item $v[p(\term_1,...,\term_n)] =
I[p][I[\term_1],...,I[\term_n]]$.
\item For every $(n,k)$-ary quantifier
$\Q$ of $L$, $v[\Q x_1,...,x_k (\psi_1,...,\psi_n)]$ should be an
element of
$\tilde{\Q}_\M
[\{\tup{v[\psi_1\{\overline{a}_1/{x}_1,...,\overline{a}_k/{x}_k\}],...,
v[\psi_n\{\overline{a}_1/{x}_1,...,\overline{a}_k/{x}_k\}]}\ |\ a_1,...,a_k\in D\}]$.

\end{enumerate}

\end{defi}

\noindent Note that in case $\Q$ is a propositional connective (for $k=0$), the function
$\tilde{\Q}_\M$ is applied to a singleton, as was explained above.\medskip

\noindent{\bf Notation.} For a set of sequents $\Sss$, we shall write $\Sss\vdash_{G}\Gamma\Rightarrow\Delta$
if a sequent $\Gamma\Rightarrow\Delta$ has a proof from $\Sss$ in
$G$.

\begin{defi}\label{fos}
Let $S=\tup{D,I}$ be an $L$-structure for an Nmatrix $\M$.
\begin{enumerate}[(1)]\item An $\M$-legal $S$-valuation $\vsg$
is {\em a model
of {a sentence} $\psi$ in $\M$}, denoted by
$S,\vsg\models_\M\psi$, if $\vsg[\psi]\in \D$.
\item Let $v$ be an $\M$-legal $S$-valuation. A sequent $\Gamma\Rightarrow\Delta$
is {\em $\M$-valid in $\tup{S,v}$} if for every $S$-substitution $\sigma$:
if $S,v\models_{\M}\sigma[\psi]$ for every $\psi\in \Gamma$, then
there is some $\varphi\in \Delta$, such that
$S,v\models_{\M}\sigma[\varphi]$.
\item A sequent $\Gamma\Rightarrow\Delta$ is $\M$-valid, denoted by $\vdash_\M\Gamma\Rightarrow\Delta$,
if for every $L$-structure $S$ and every $\M$-legal $S$-valuation $v$, $\Gamma\Rightarrow\Delta$
is $\M$-valid in $\tup{S,v}$.

\item For a set of sequents $\Sss$,
$\Sss\vdash_{\M}\Gamma\Rightarrow\Delta$ if for every $L$-structure $S$ and every $\M$-legal $S$-valuation $v$:
whenever the sequents of $\Sss$ are $\M$-valid in $\tup{S,v}$, $\Gamma\Rightarrow\Delta$
is also $\M$-valid in $\tup{S,v}$.

\end{enumerate}
\end{defi}

\begin{defi} A system $G$ is {\em strongly sound}\footnote{A more general definition would be without
the restriction concerning the closure of $\Sss$ under substitution. However,
in this case we would need to add substitution as a structural rule to canonical calculi.}
for an Nmatrix $\M$ if for every set $\Sss$ of sequents closed under
substitution: $\Sss\vdash_G\Gamma\Rightarrow\Delta$
entails $\Sss\vdash_\M\Gamma\Rightarrow\Delta$. A system $G$ is {\em strongly complete}
for an Nmatrix $\M$ if for every set $\Sss$ of sequents closed under
substitution: $\Sss\vdash_\M\Gamma\Rightarrow\Delta$
entails $\Sss\vdash_G\Gamma\Rightarrow\Delta$. An Nmatrix $\M$ is
{\em strongly characteristic} for $G$ if $G$ is strongly sound and
strongly complete for $\M$.
\end{defi}

Note that since the empty set of sequents is closed under
substitutions, strong soundness implies (weak) soundness\footnote{A
  system $G$ is (weakly) sound for an Nmatrix $\M$ if
  $\vdash_G\Gamma\Rightarrow\Delta$ entails
  $\vdash_\M\Gamma\Rightarrow\Delta$.}. A similar remark applies to
completeness and a characteristic Nmatrix.

%

\subsection{Semantics for simplified languages \texorpdfstring{$\lmon$}{lmon}}
In addition to $L$-structures for languages with $(n,k)$-ary
quantifiers, we also use $\lmon$-structures for the simplified
languages $\lmon$, used for formulating the canonical rules. To
make the distinction clearer, we shall use the metavariable $S$
for the former and $\C$ for the latter. Since the formulas of
$\lmon$ are always atomic, the specific 2Nmatrix for which $\C$ is
defined is immaterial, and can be omitted. We may even speak of
classical validity of sequents over $\lmon$. Thus henceforth instead of
speaking of $\M$-validity of a set of clauses $\Theta$ over
$\lmon$, we will speak simply of validity.

Next we define the notion of a {\em distribution} of
$\lmon$-structures.

\begin{defi}Let
$\C=\tup{D,I}$ be a structure for $\lmon$. $Dist_\C$,
the distribution of $\C$ is defined as follows:
\[Dist_\C = \{\tup{I[p_1][a_1,...,a_k],...,I[p_n][a_1,...,a_k]}\ |\ a_1,...,a_k\in D\}\]
We say that an $\lmon$-structure $\C$ is $\E$-characteristic if
$Dist_\C = \E$.
\end{defi}
Note that the distribution of an ${\mathcal{L}}^n_0$-structure $\C$ is
$Dist_\C = \{\tup{I[p_1],...,I[p_n]}\}$ and so it is always
a singleton. Furthermore, the validity of a set of clauses over
${\mathcal{L}}^n_0$ can be reduced to propositional satisfiability as
stated in the following lemma which can be easily proved: \remove{\begin{lem}Let $\Theta$
be a set of clauses over $\lmon$. Let Let $\E\subseteq
\{t,f\}^n$. The question whether $\Theta$ is
satisfiable\footnote{A clause $\Gamma\Rightarrow\Delta$ is
satisfiable in a $\lmon$-structure $\C$ for an 2Nmatrix $\M$ if
for every substitution $\sigma$ either there exists some
$p_i(\term)\in \Gamma$, such that $I[p_i][I[\term]]=f$, or there
is some $p_i(\term)\in \Delta$, such that $I[p_i][I[\term]]=f$.
Note that since only atomic formulas are involved, it is not
necessary to specify the 2Nmatrix. } by an $\E$-characteristic
structure is decidable.
\end{lem}
}

\begin{lem}\label{propo}Let $\C$ be a
${\mathcal{L}}^n_0$-structure. Assume that $Dist_\C = \{\tup{s_1,...,s_n}\}$ for some $s_1,...,s_n\in \{t,f\}$.
Let $v_{Dist_\C}$ be
any propositional valuation satisfying $v[p_i]=s_i$ for every $1\leq i \leq n$. A set of
clauses $\Theta$ is valid in $\C$ iff $v_{Dist_\C}$ propositionally
satisfies $\Theta$.
\end{lem}
Now we turn to the case $k=1$. In this case it is convenient to
define a special kind of ${\mathcal{L}}^n_1$-structures which we call
{\em canonical} structures. These structures are sufficient to reflect
the behavior of all possible ${\mathcal{L}}^n_1$-structures.
\begin{defi}Let $\E\in P^+(\{t,f\}^n)$.
A ${\mathcal{L}}^n_1$-structure $\C = \tup{D,I}$ is $\E$-canonical if
$D = \E$ and for every $b=\tup{s_1,...,s_n}\in D$ and every $1\leq
i \leq n$:  $I[p_i][b]=s_i$.
\end{defi}

Clearly, every $\E$-canonical ${\mathcal{L}}^n_1$-structure is
$\E$-characteristic.
\begin{lem}\label{canstr}Let $\Theta$ be a set of clauses over ${\mathcal{L}}^n_1$, which is
valid in some structure $\C=\tup{D,I}$. Then there exists a
$Dist_\C$-canonical structure $\C'$ in which $\Theta$ is valid.
\end{lem}
\proof Suppose that $\Theta$ is valid in a structure $\C =
\tup{D,I}$. Define the ${\mathcal{L}}^n_1$-structure $\C'=\tup{I',D'}$ as follows:
\begin{enumerate}[$\bullet$]
\item $D' = Dist_\C$.
\item $I'[c] =
\tup{I[p_1][I[c]],...,I[p_n][I[c]]}$ for every constant $c$
occurring in $\Theta$.

\item For every $1\leq i \leq
n$: $I'[p_i][\tup{s_1,...,s_n}]=t$ iff $s_i=t$.

\end{enumerate}
Clearly, $\C'$ is $Dist_\C$-canonical. It is easy to verify
that $\Theta$ is valid in $\C'$. \qed

\begin{cor}\label{decf}Let $\E\in P^+(\{t,f\}^n)$.
For a finite set of clauses $\Theta$ over ${\mathcal{L}}^n_1$, the question
whether $\Theta$ is valid in a $\E$-characteristic structure is
decidable.
\end{cor}
\proof Follows from Lemma \ref{canstr} and the fact that for any $\E\in P^+(\{t,f\}^n)$,
there are finitely many $\E$-canonical structures to check.
\section{Canonical systems with \texorpdfstring{$(n,k)$}{(n,k)}-ary
  quantifiers for \texorpdfstring{$k\in \{0,1\}$}{k in {0,1}}}
\noindent Now we turn to the class of canonical systems with $(n,k)$-ary quantifiers for
the case of $k\in \{0,1\}$ and $n\geq 1$. Henceforth, unless stated
otherwise, we assume that $k\in \{0,1\}$.
\subsection{Semantics for canonical systems for \texorpdfstring{$k\in
    \{0,1\}$}{k in {0,1}}} 
\indent In this section we explore the connection between the
coherence of a canonical calculus $G$, the existence for it of a
strongly characteristic 2Nmatrix, and strong cut-elimination (in
a sense explained below.) We start by defining the notion of {\em suitability} for $G$.
\begin{defi}\label{suit}{\bf (Suitability for $G$)} Let $G$ be a canonical calculus over $L$.
A 2Nmatrix $\M$ is {\em suitable} for $G$ if for every $(n,k)$-ary
canonical rule $\Theta/\Q(s)$ of $G$ (where
$s\in \{t,f\}$), it holds
that for every $\lmon$-structure $\C$ in which $\Theta$ is valid:
$\tilde{\Q}_\M[Dist_\C] = \{s\}$.
\end{defi}

\begin{thm}\label{sound}Let $G$ be a canonical calculus and $\M$ - a
2Nmatrix suitable for $G$. Then $G$ is strongly sound for $\M$.
\end{thm}
\proof see Appendix \ref{app}.\medskip

Now we come to the construction of a characteristic
2Nmatrix for every coherent canonical calculus.
\begin{defi}\label{charm}Let $G$ be a coherent canonical calculus.
The Nmatrix $\M_G$ for $L$ is defined as
follows for every $(n,k)$-ary quantifier $\Q$ of $L$, every $s\in
\{t,f\}$ and every $\E\in P^+(\{t,f\}^n)$:
\[\tilde{{\Q}}_{\M_G}[\E] = \begin{cases}
  \{s\} & \mathit{if\ \Theta/\Q(s)\in G\ and}\\
  & \mathit{\Theta\ is\ valid\ in\  some\
  \E\!-\!canonical\ \lmon-structure} \\
  \{t,f\} & {otherwise}
\end{cases}\]

\end{defi}
\noindent First of all, note that by corollary \ref{decf}, the above definition is
constructive. Next, let us show that $\M_G$ is well-defined. Assume by
contradiction that there are two dual rules $\Theta_1/\Rightarrow A$
and $\Theta_2/A\Rightarrow$, such that both $\Theta_1$ and
$\Theta_2$ are valid in some $\E$-canonical structures
$\C_1,\C_2$ respectively. Obtain
$\Theta'_2$ from $\Theta_2$ by renaming of constants and variables
which occur in $\Theta_1$. Then clearly $\Theta'_2$ is also valid in some
$\E$-canonical structure $\C_3$.
If $k=0$, by
Lemma \ref{propo}, the set of clauses $\Theta_1\cup\Theta'_2$ is
satisfiable by a (classical) propositional valuation $v_\E$ and is thus classically
consistent, in contradiction to the
coherence of $G$ (see defn. \ref{cohdefn}).\\
Otherwise, $k=1$. The only difference between different $\E$-canonical
structures is in the interpretation of constants, and since the
sets of constants occurring in $\Theta_1$ and $\Theta'_2$ are
disjoint, an $\E$-canonical structure $\C'=\tup{D',I'}$ (for the
extended language containing the constants of both $\Theta_1$ and
$\Theta_2$) can be constructed, in which $\Theta_1\cup\Theta'_2$
are valid. Thus the set $\Theta_1\cup\Theta'_2={\sf Rnm}(\Theta_1\cup\Theta_2)$ is classically
consistent, in contradiction to the coherence of $G$.\\ \ \\
\noindent {\bf Remark:} The construction of $\M_G$ above is much simpler
than the constructions carried out in \cite{AvrLev01,Zam_SL}:
a canonical calculus there is first transformed
into an equivalent normal form calculus, which is then used to
construct the characteristic Nmatrix. The idea is to transform the
calculus so that each rule dictates the interpretation for only
one $\E$. However, the above definitions show that the
transformation into normal form is actually not necessary and we
can construct $\M_G$ directly from $G$.

Next we demonstrate the construction of a characteristic 2Nmatrix for some
coherent canonical calculi.
\begin{exas}\label{ex2}
\begin{enumerate}[(1)]
\item
It is easy to see that for any canonical coherent calculus $G$ including the
standard (1,1)-ary rules for $\forall$ and $\exists$ from Example \ref{ex1}-2:
\[\tilde{\forall}_{\M_G}[\{t,f\}] = \tilde{\forall}_{\M_G}[\{f\}] =
\tilde{\exists}_{\M_G}[\{f\}]=\{f\}\]
\[\tilde{\forall}_{\M_G}[\{t\}] = \tilde{\exists}_{\M_G}[\{t,f\}] =
\tilde{\exists}_{\M_G}[\{t\}]=\{t\}\]

\item Consider the canonical calculus $G'$ consisting of the
following three $(1,2)$-ary rules from
Example \ref{ex1}-3:
\[\{p_1(v_1)\Rightarrow
p_2(v_1)\}/\Rightarrow\overline{\forall} v_1\ (p_1(v_1),p_2(v_1))
\]
\[\{ p_2(c)\Rightarrow\ ,\ \Rightarrow p_1(c)\}/\overline{\forall} v_1 (p_1(v_1),p_2(v_1))\Rightarrow\]
\[  \{\Rightarrow p_1(c)\ ,\ \Rightarrow
p_2(c)\}/\Rightarrow\overline{\exists} v_1 (p_1(v_1),p_2(v_1))\]
$G'$ is obviously coherent. The 2Nmatrix $\M_{G'}$ is defined as follows for every $H\in P^+(\{t,f\}^2)$:
\[\tilde{\overline{\forall}}[H] = \begin{cases}
  \{t\} & \mathit{if\ \tup{t,f}\nin H}\\
  \{f\} & {otherwise}
\end{cases}\ \ \ \ \ \ \tilde{\overline{\exists}}[H] = \begin{cases}
  \{t\} & \mathit{if\ \tup{t,t}\in H}\\
  \{t,f\} & {otherwise}
\end{cases}\]
The first rule dictates the condition that $\overline{\forall}[H]=\{t\}$ for
the case of $\tup{t,f}\nin H$. The second rule dictates the
condition that
$\overline{\forall}[H]=\{f\}$ for the case that $\tup{t,f}\in H$.
Since $G'$ is coherent, these conditions are non-contradictory.
The third rule dictates the condition that $\overline{\exists}[H]=\{t\}$
in the case that $\tup{t,t}\in H$. There is no rule which dictates
conditions for the case of $\tup{t,t}\nin H$, and so the
interpretation in this case is non-deterministic.

\item Consider the canonical calculus $G''$ consisting of the
following $(1,3)$-ary rule:
\[\{p_2(v_1),p_3(v_1)\Rightarrow\}/\Q v_1 (p_1(v_1),p_2(v_1),p_3(v_1))\Rightarrow\]
Of course, $G''$ is coherent. The 2Nmatrix $\M_{G''}$ is defined as follows for every $H\in P^+(\{t,f\}^2)$:
\[\tilde{\overline{\forall}}[H] = \begin{cases}
  \{f\} & \mathit{if\ H\subseteq \{\tup{t,t,f},\tup{t,f,t},\tup{t,f,f},\tup{f,t,f},\tup{f,f,t},\tup{f,f,f}\}}\\
  \{t,f\} & \mathit{if\ \tup{f,t,t}\in H\ or\ \tup{t,t,t}\in H}
\end{cases}\]
\end{enumerate}
\end{exas}


\noindent Now we come to the main theorem, establishing a connection
between the coherence of a canonical calculus $G$, the existence of a
strongly characteristic 2Nmatrix for $G$ and {\em strong
  cut-elimination} in $G$ in the sense of \cite{Avr93}.

\begin{defi}Let $G$ be a canonical calculus and let $\Sss$ be a set of sequents closed under
substitution. A proof $P$ of $\Gamma\Rightarrow\Delta$ from $\Sss$ in
$G$ is {\em simple} if all cuts in $P$ are on formulas
from $\Sss$.

\end{defi}

\begin{defi}A calculus $G$ admits {\em strong cut-elimination}\footnote{\cite{Avr93} does not assume
that $\Sss$ is closed under substitution. Instead, a structural substitution rule is added and the allowed cuts
are on substitution instances of formulas from $\Sss$.} if for
every set of sequents $\Sss$ closed under substitution and every sequent
$\Gamma\Rightarrow\Delta$, such that
$\Sss\cup\{\Gamma\Rightarrow\Delta\}$ satisfies the free-variable
condition\footnote{See section \ref{prelim}.}:
if $\Sss\vdash_G\Gamma\Rightarrow\Delta$, then
$\Gamma\Rightarrow\Delta$ has a simple proof in $G$.

\end{defi}

Note that strong cut-elimination implies standard cut-elimination
(which corresponds to the case of an empty set $\Sss$).

\begin{thm}\label{main}Let $G$ be a canonical calculus. Then the following
statements concerning $G$ are equivalent:
\begin{enumerate}[\em(1)]
\item $G$ is coherent.

\item $G$ has a strongly characteristic 2Nmatrix.

\item $G$ admits strong cut-elimination.

\end{enumerate}
\end{thm}

\proof\noindent First we prove that (2) implies
  (1). Suppose that $G$ has a strongly characteristic 2Nmatrix
$\M$. Assume by contradiction that $G$ is not coherent. Then there
exist two dual $(n,k)$-ary rules $R_1=\Theta_1/\Rightarrow A$ and
$R_2=\Theta_2/A\Rightarrow$ in $G$, such that ${\sf
  Rnm}(\Theta_1\cup\Theta_2)$ is classically consistent. Suppose that
$k=1$. Then $A = \Q v_1 (p_1(v_1),...,p_n(v_1))$. Recall that ${\sf
  Rnm}(\Theta_1\cup\Theta_2) = \Theta_1\cup\Theta'_2$, where
$\Theta'_2$ is obtained from $\Theta_2$ by renaming constants and
variables that occur also in $\Theta_1$ (see defn. \ref{renaming}).
For simplicity\footnote{This assumption is not necessary and is used
  only for simplification of presentation, since we can instantiate
  the constants by any $L$-terms.} we assume that the fresh constants
used for renaming are all in $L$.  Let $\Theta_1 =
\{\Sigma^1_j\Rightarrow\Pi^1_j\}_{1\leq j \leq m}$ and $\Theta'_2 =
\{\Sigma^2_j\Rightarrow\Pi^2_j\}_{1\leq j \leq r}$. Since
$\Theta_1\cup\Theta'_2$ is classically consistent, there exists an
$\lmon$-structure $\C=\tup{D,I}$, in which both $\Theta_1$ and
$\Theta'_2$ are valid. Recall that we also assume that $\lmon$ is a
subset of $L$\footnote{This assumption is again not essential for the
  proof, but it simplifies the presentation.} and so the following are
applications of $R_1$ and $R_2$ respectively:
\[\infer[]{\Rightarrow \Q v_1 (p_1(v_1),...,p_n(v_1))}{\{\Sigma^1_j\Rightarrow\Pi^1_j\}_{1\leq j \leq m}}
\ \ \ \ \infer[]{ \Q v_1 (p_1(v_1),...,p_n(v_1))\Rightarrow}{\{\Sigma^2_j\
\Rightarrow\Pi^2_j\}_{1\leq j \leq m}}
\]
Let $S$ be any extension of $\C$ to $L$ and $v$ - any $\M$-legal $S$-valuation.
It is easy to see that
the premises of the applications above are $\M$-valid in $\tup{S,v}$ (since the premises contain atomic formulas).
Since $G$ is strongly sound for $\M$, both $\Rightarrow \Q v_1 (p_1(v_1),...,p_n(v_1))$
and $ \Q v_1 (p_1(v_1),...,p_n(v_1))\Rightarrow$ should also be
$\M$-valid in $\tup{S,v}$, which is of course impossible.
The proof for the case of $k=0$ is simpler and is left to the
reader.

Next, we prove that (3) implies (1). Let $G$ be a
canonical calculus which admits strong cut-elimination. Suppose by
contradiction that $G$ is not coherent. Then there are two dual rules
of $G$: $\Theta_1/\Rightarrow A$ and $\Theta_2/A\Rightarrow$, such
that ${\sf Rnm}(\Theta_1\cup\Theta_2)$ is classically consistent. Let
$\Theta$ be the minimal set of clauses, such that ${\sf
  Rnm}(\Theta_1\cup\Theta_2)\subseteq \Theta$ and $\Theta$ is closed
under substitutions. $\Theta\cup\{\Rightarrow\}$ satisfy the
free-variable condition, since only atomic formulas are involved and
no variables are bound there. It is easy to see that $\Theta\vdash_G
\Rightarrow A$ and $\Theta\vdash_G A\Rightarrow$. By using cut,
$\Theta\vdash_G\Rightarrow$.  But $\Rightarrow$ has no simple proof in
$G$ from $\Theta$ (since ${\sf Rnm}(\Theta_1\cup\Theta_2)$ is
consistent and $\Theta$ is its closure under substitutions), in
contradiction to the fact that $G$ admits strong cut-elimination.

To show that (1) implies both (2) and (3), we need the following
proposition:

\begin{prop}\label{mainprop}Let $G$ be a coherent calculus. Let $\Sss$ 
be a set of sequents closed under substitution and
$\Gamma\Rightarrow\Delta$ - a sequent, such that
$\Sss\cup\{\Gamma\Rightarrow\Delta\}$ satisfies the free-variable
condition. If $\Gamma\Rightarrow\Delta$ has no simple proof from
$\Sss$ in $G$, then $\Sss{\not\vdash_\M}\Gamma\Rightarrow\Delta$.

\end{prop}
\proof see Appendix \ref{app}.

To prove that (1) implies (2), suppose that $G$ is coherent. Let us
show that $\M_G$ is a strongly characteristic 2Nmatrix for $G$. By
definition of $\M_G$, it is suitable for $G$ (see
defn. \ref{suit}). By theorem \ref{sound}, $G$ is strongly sound for
$\M_G$.\\ For strong completeness, let $\Sss$ be a set of sequents
closed under substitution. Suppose that a sequent
$\Gamma\Rightarrow\Delta$ has no proof from $\Sss$ in $G$. If
$\Sss\cup\{\Gamma\Rightarrow\Delta\}$ does not satisfy the
free-variable condition, obtain $\Sss'\cup
\{\Gamma'\Rightarrow\Delta'\}$ by renaming the bound variables, so
that $\Sss'\cup\{\Gamma'\Rightarrow\Delta'\}$ satisfies the condition
(otherwise, take $\Gamma'\Rightarrow\Delta'$ and $\Sss'$ to be
$\Gamma\Rightarrow\Delta$ and $\Sss$ respectively). Then
$\Gamma'\Rightarrow\Delta'$ has no proof from $\Sss'$ in $G$
(otherwise we could obtain a proof of $\Gamma\Rightarrow\Delta$ from
$\Sss$ by using cuts on logical axioms), and so it also has no simple
proof from $\Sss'$ in $G$. By Proposition \ref{mainprop},
$\Sss'{\not\vdash_\M}\Gamma'\Rightarrow\Delta'$.  That is, there is an
$L$-structure $S$ and an $\M$-legal valuation $v$, such that the
sequents in $\Sss'$ are $\M$-valid in $\tup{S,v}$, while
$\Gamma'\Rightarrow\Delta'$ is not. Since $v$ respects the
$\equiv_\alpha$-relation, the sequents of $\Sss$ are also $\M$-valid
in $\tup{S,v}$, while $\Gamma\Rightarrow\Delta$ is not. And so
$\Sss{\not\vdash_\M}\Gamma\Rightarrow\Delta$. We have shown that $G$
is strongly complete (and strongly sound) for $\M_G$. Thus $\M_G$ is a
strongly characteristic 2Nmatrix for $G$.

Finally, we prove that (1) implies (3). Let $G$ be a coherent
calculus. Let $\Sss$ be a set of sequents closed under substitution,
and let $\Gamma\Rightarrow\Delta$ be a sequent, such that $\Sss\cup
\{\Gamma\Rightarrow\Delta\}$ satisfies the free-variable
condition. Suppose that $\Sss\vdash_G\Gamma\Rightarrow\Delta$. We have
already shown above that $\M_G$ is a strongly characteristic 2Nmatrix
for $G$. Thus $\Sss\vdash_{\M}\Gamma\Rightarrow\Delta$, and by
Proposition \ref{mainprop}, $\Gamma\Rightarrow\Delta$ has a simple
proof from $\Sss$ in $G$. Thus $G$ admits strong cut-elimination. \qed\medskip

\noindent {\bf Remark.} 
At this point it should be noted that the
renaming of clashing constants in the definition of
coherence (see defn. \ref{cohdefn}) is crucial. Consider, for instance, a canonical calculus $G$
consisting of the introduction rules $\{p_1(c)\Rightarrow\ ;\ \Rightarrow
p_1(c')\}/\Rightarrow {{\Q}} v_1\  p_1(v_1)$ and
$\{p_1(c'')\Rightarrow\ ;\
\Rightarrow p_1(c)\}/{{\Q}} v_1\  p(v_1)\Rightarrow$ for a (1,1)-ary quantifier $\Q$.
Without renaming of clashing constants, we would conclude
that the set $\{p_1(c)\Rightarrow\ ;\ \Rightarrow
p_1(c')\ ;\ p_1(c'')\Rightarrow,
\Rightarrow p_1(c)\}$ is classically inconsistent. However, $G$
obviously has no
strongly characteristic 2Nmatrix, since the rules dictate contradicting
requirements for $\tilde{\Q}[\{t,f\}]$. But if we perform renaming first, obtaining the
set ${\sf Rnm}(\Theta_1\cup\Theta_2)=\{p_1(c)\Rightarrow\ ,\ \Rightarrow
p_1(c')\ ,\ p_1(c'')\Rightarrow,
\Rightarrow p_1(c''')\}$, we shall see that ${\sf
Rnm}(\Theta_1\cup\Theta_2)$ is classically consistent and so $G$ is not coherent. Hence, by
the above theorem, $G$ has no strongly characteristic
2Nmatrix.\medskip

\remove{
: Let $G$ have a strongly characteristic 2Nmatrix $\M$. Suppose by
contradiction that $G$ is not coherent. Then there are two $(n,k)$-ary rules of $G$: $\Theta_1/\Rightarrow \Q z^k(p_1(z^k),...,p_n(z^k))$
and $\Theta_2/\Q z^k(p_1(z^k),...,p_n(z^k))\Rightarrow$ (where $k\in \{0,1\}$), such that
$\Theta_1\cup\Theta_2$ is classically consistent. Then there is some $\lmon$-structure
$\C$, in which both $\Theta_1$ and $\Theta_2$ are valid (in particular, they are $\M$-valid).
Since $\M$ is strongly characteristic,
it is strongly sound for $G$, and so both the conclusions $\Q z^k(p_1(z^k),...,p_n(z^k))\Rightarrow$
and $\Rightarrow\Q z^k(p_1(z^k),...,p_n(z^k))$ are $\M$-valid at the same time, which is of course impossible.\\
(2$\Rightarrow$1): Let $G$ be a coherent canonical calculus. Then by theorem \ref{maint},
$\M_G$ is a strongly characteristic 2Nmatrix for $G$.}
\begin{cor}The existence of a strongly characteristic 2Nmatrix for a canonical calculus
$G$ is decidable.
\end{cor}
\proof By theorem \ref{main}, the question whether $G$ has a strongly characteristic 2Nmatrix is
equivalent to the question whether $G$ is coherent, and this, by
Proposition \ref{decidability}, is decidable.\\ \ \\
{\bf Remark:} The above results are related to the results in
\cite{Ciabnk}, where a general class of sequent calculi with
$(n,k)$-ary quantifiers and a (not necessarily standard) set of structural rules called {\em standard}
calculi are defined. A canonical calculus is a
particular instance of a standard calculus which includes all of the standard structural rules.
\cite{Ciabnk} formulate syntactic necessary and sufficient
conditions
for a slightly generalized version of cut-elimination with non-logical
axioms. Unlike in this paper, the non-logical axioms must
consist of {\em atomic} formulas (and must be closed under cuts and
substitutions). But the results of \cite{Ciabnk} apply to a much wider class of
calculi (since different combinations of
structural rules are allowed). In addition, a constructive modular cut-elimination
procedure is provided. The
{reductivity} condition of \cite{Ciabnk} can be shown to be equivalent to
our coherence criterion in the context of canonical systems\footnote{We wish to thank
Agata Ciabattoni for pointing out these facts to us in a personal correspondence.} .

\subsection{Coherence and standard cut-elimination}

In the previous subsection we have studied the connection between
coherence and strong cut-elimination. In this subsection we focus on
standard cut-elimination in canonical calculi. It easily follows from
theorem \ref{main} that coherence implies cut-elimination:

\begin{cor}Let $G$ be a canonical calculus. If $G$ is coherent,
then for every sequent $\Gamma\Rightarrow\Delta$ satisfying the free-variable
condition: if $\Gamma\Rightarrow\Delta$ is provable in $G$, then
it has a cut-free proof in $G$.
\end{cor}

Thus coherence is a sufficient condition for cut-elimination in a
canonical calculus. In the more restricted canonical systems of
\cite{AvrLev01,Zam_SL} it also is a necessary condition. However,
things get more complicated with the more general canonical
rules studied in this paper. 

\begin{exa}\label{notcoh} Consider, for instance, the following canonical calculus $G_0$ consisting
of the following two inference rules: $\Theta_1/\Rightarrow \Q v_1
(p_1(v_1),p_2(v_1))$ and $\Theta_2/\Q v_1
(p_1(v_1),p_2(v_1))\Rightarrow$, where:
\[\Theta_1\!=\!\Theta_2\!=\!\{p_1(v_1)\Rightarrow p_2(v_1)\ ;
\Rightarrow p_1(c_1)\ ;\Rightarrow p_2(c_1)\ ;p_1(c_2)\Rightarrow\ ;p_2(c_2)\Rightarrow\ ;p_1(c_3)
\Rightarrow\ ;\Rightarrow p_2(c_3)
\}\]

Clearly, $G_0$ is not coherent. We now sketch a proof that the only
sequents provable in $G_0$ are logical axioms. This immediately
implies that $G_0$ admits cut-elimination.

To prove this it suffices to show that for every rule of $G_0$: if its
premises are logical axioms, then its conclusion is a logical axiom.
Suppose by contradiction that we can apply one of the rules on logical
axioms and obtain a conclusion which is not a logical axiom. Suppose,
without loss of generality, that it is the first rule. Then the
application would be of the form:

\[\infer[]{\Gamma\Rightarrow\Q w (\chi[p_1],\chi[p_2]),\Delta}
{\Gamma,\chi[p_1]\{\chi[v_1]/w\}\Rightarrow
  \Delta,\chi[p_2]\{\chi[v_1]/w\!\!&\!\!\dots\!\!
&\!\! \Gamma\Rightarrow \chi[p_1]\{\chi[c_1]/w\},\Delta & \Gamma\Rightarrow \chi[p_2]\{\chi[c_1]/w\},\Delta}
\]
Since the proved sequent is not a logical axiom, (*) there are no $A\in \Gamma$
and $B\in \Delta$, such that $A\equiv_\alpha B$. Moreover, since $\Gamma,\chi[p_1]\{\chi[v_1]/w\}\Rightarrow \Delta,\chi[p_2]\{\chi[y]/w\}$
is a logical axiom, either (i) there is some $C\in \Delta$, such that
$C\equiv_\alpha \chi[p_1]\{\chi[v_1]/w\}$,
(ii) there is some $C\in \Gamma$, such that $C\equiv_\alpha \chi[p_2]\{\chi[v_1]/w\}$, or (iii)
$\chi[p_1](\chi[v_1]/w)\equiv_\alpha\chi[p_2]\{\chi[v_1]/w\}$. Suppose (i)
holds, i.e. there is some some $C\in \Delta$, such that
$C\equiv_\alpha \chi[p_1]\{\chi[v_1]/w\}$.
Then since $\chi[v_1]$ cannot occur free in $\Delta$, $w\nin
Fv[C]$, and so $w\nin Fv[\chi[p_1]]$. Hence,
$\chi[p_1]\{\chi[c_1]/w\}=\chi[p_1]\{\chi[v_1]/w\}=\chi[p_1]$.
Now since $\Gamma\Rightarrow \chi[p_1]\{\chi[c_1]/w\},\Delta$ is a
logical axiom, and due to (*), there is some $D\in \Gamma$, such
that $D\equiv_\alpha \chi[p_1]\{\chi[c_1]/w\}$. But since
$\chi[p_1]\{\chi[c_1]/w\}=\chi[p_1]\{\chi[v_1]/w\}$, $C\equiv_\alpha
D$, $C\in \Delta$ and $D\in \Gamma$, in contradiction to (*). The
case (ii) is treated similarly using the constant $c_2$. The case
(iii) is handled using the constant $c_3$.

Thus, only logical axioms are provable in $G_0$ and so it admits
standard cut-elimination, although it is not coherent.\end{exa}

Hence coherence is not a necessary condition for cut-elimination in
general. However, below we characterize a more restricted subclass of
canonical systems, for which this property does hold.

\begin{defi}\label{simple}A canonical calculus $G$ is {\em simple} if for every
two dual $(n,k)$-ary canonical rules $\Theta_1/\Rightarrow A$ and
$\Theta_2/A\Rightarrow$ one of the following properties holds:
\begin{enumerate}[(1)]

\item $k=0$, i.e. $\Theta_1/\Rightarrow A$ and
$\Theta_2/A\Rightarrow$ are propositional rules.

\item $k=1$ and one of the following holds for each variable $y$ occurring in ${\sf Rnm}(\Theta_1\cup\Theta_2)$:
\begin{enumerate}[$\bullet$]
\item There is at most one $1\leq i\leq n$, such that
$y$ occurs in $p_i(y)$
in ${\sf Rnm}(\Theta_1\cup\Theta_2)$ and there is at most
one constant $c$, such that $p_i(c)$ also occurs in
${\sf Rnm}(\Theta_1\cup\Theta_2)$.

\item There are two different $1\leq i,j\leq n$, such that
$y$ occurs in $p_i(y)$ and $p_j(y)$
in ${\sf Rnm}(\Theta_1\cup\Theta_2)$ and for every constant $c$,
there is no such $1\leq k \leq n$, that both $p_k(y)$ and $p_k(c)$
occur in ${\sf Rnm}(\Theta_1\cup\Theta_2)$.

\end{enumerate}
\end{enumerate}

\end{defi}

\begin{exas}
\begin{enumerate}[(1)]
\item All the canonical calculi from examples \ref{ex1} are
simple.

\item Consider the canonical calculus $G_1$, consisting of
the following two rules for a $(3,1)$-ary quantifier $\Q_1$:
\[\{p_1(v_1)\Rightarrow\ ;\ p_1(c),p_2(c)\Rightarrow\ \}/\Rightarrow \Q_1 v_1(p_1(v_1),p_2(v_1),
p_3(v_1))\]
\[\{\Rightarrow p_1(v_1)\ ;\ \Rightarrow p_2(e)\}/\Q_1v_1(p_1(v_1),p_2(v_1),p_3(v_1))\Rightarrow\]
It is easy to see that $G_1$ is a simple coherent calculus.

\item If we modify the first rule of $G_1$ as follows:
\[\{p_1(v_1)\Rightarrow\ ;\ p_1(c),p_2(c)\Rightarrow\ ;
\ p_1(d)\Rightarrow p_3(d)\}/\Rightarrow \Q_1 v_1(p_1(v_1),p_2(v_1),p_3(v_1))\]
the resulting calculus is not simple, since both $p_1(c)$ and
$p_1(d)$ occur in the premises of the rule, together with
$p_1(v_1)$.

\item The calculus $G_0$ from example \ref{notcoh} is not simple, since
for instance $p_1(v_1)$, $p_1(c_1)$ and $p_1(c_5)$ occur in the premises (after renaming).

\end{enumerate}
\end{exas}

\begin{prop}\label{simple-prop}If a simple canonical calculus $G$ admits
cut-elimination, then it is coherent.
\end{prop}
\proof see Appendix \ref{app}.

\section{Summary and further research}

In this paper we have considerably extended the characterization of
canonical calculi of \cite{AvrLev01,Zam_SL} to $(n,k)$-ary
quantifiers. Focusing on the case of $k\in \{0,1\}$, we have shown
that the following statements concerning a canonical calculus $G$ are
equivalent: (i) $G$ is coherent, (ii) $G$ has a strongly
characteristic 2Nmatrix, and (iii) $G$ admits strong
cut-elimination. We have also shown that coherence is not a necessary
condition for standard cut-elimination, and characterized a subclass
of canonical systems called {\em simple} calculi, for which this
property does hold.

In addition to these proof-theoretical results for a natural type of
multiple conclusion Gentzen-type systems with $(n,1)$-ary and
$(n,0)$-ary quantifiers, this work also provides further evidence for
the thesis that the meaning of a logical constant is given by its
introduction (and ``elimination'') rules . We have shown that at least
in the framework of multiple-conclusion consequence relations, any
``reasonable'' set of canonical quantificational rules completely
determines the semantics of the quantifier. \\ \indent This paper also
demonstrates the important role of the semantic framework of Nmatrices
(\cite{AvrLev01,Zam_ISMVL}), which substantially contributes to the
understanding of the connection between syntactic rules and semantic
interpretations of quantifiers. Due to the modularity of the
framework, we were able to detect the semantic effect of each of the
canonical rules, which of course is not possible using deterministic
matrices.

Some of the most immediate research directions are as follows. In the
case of $k\in \{0,1\}$, we still need to characterize the most general
subclass of canonical calculi, for which coherence is both a necessary
and sufficient condition for standard cut-elimination (it is not clear
whether the characterization of simple calculi can be further
extended).

Extending these results to the case of $k>1$ might lead to new
insights on Henkin quantifiers and other important generalized
quantifiers.  However, even for the simplest case of $(1,2)$-ary
quantifiers the extension is far from straightforward. Consider, for
instance, the calculus $G$, consisting of the following two (1,2)-ary
rules:
\[\{p(c,x)\Rightarrow\}/\Rightarrow \Q z_1z_2p(z_1,z_2)\ \ \ \{\Rightarrow p(y,d)\}/ \Q z_1z_2p(z_1,z_2)\Rightarrow\]
$G$ is coherent, but it is easy to see that $\M_G$ is not well-defined in this case. And
even if a 2Nmatrix $\M$ suitable for $G$
does exist, it is not necessarily sound for $G$. It is clear that the distributional interpretation of quantifiers
is no longer adequate for the case of $k>1$, since it cannot capture any kind of dependencies between
elements of the domain. Thus a more general interpretation of
quantifiers is needed.\\
\indent Another important research direction is extending canonical
systems with equality. This will allow us to treat counting
$(n,k)$-ary quantifiers, like ``there are at most two elements $a,b$, such
that $p(a,b)$ holds". Clearly, equality must be incorporated also into the
representation language $\lmon$. Standard and strong cut-elimination and its
connection to the coherence of canonical systems are yet to be
investigated for canonical systems with equality.


\section*{Acknowledgement}
\noindent This research was supported by the {\em Israel Science Foundation}
founded by the Israel Academy of
Sciences and Humanities (grant No 809/06).

\appendix\section{Proofs of selected propositions}\label{app}

\noindent {\bf Proof of Theorem \ref{sound}}: Suppose that $\M$ is
suitable for $G$. Let $S=\tup{D,I}$ be some $L$-structure and $v$ - an
$\M$-legal $S$-valuation.  Let $\Sss$ be any set of sequents closed
under substitution. We will show that if the sequents of $\Sss$ are
$\M$-valid in $\tup{S,v}$, then any sequent provable from $\Sss$ in
$G$ is $\M$-valid in $\tup{S,v}$. Obviously, the axioms of $G$ are
$\M$-valid, and the structural rules, including cut, are strongly
sound. It remains to show that for every application of a canonical
rule $R$ of $G$: if the premises of $R$ are $\M$-valid in $\tup{S,v}$,
then its conclusion is $\M$-valid in $\tup{S,v}$. We will show this
for the case of $k=1$, leaving the easier case of $k=0$ to the reader.

Let $R$ be an $(n,1)$-ary rule of $G$:
\[R=\Theta_R/\Q
v_1 (p_1(v_1),...,p_n(v_1))^{-r}\Rightarrow \Q
v_1 (p_1(v_1),...,p_n(v_1))^{r}\] where $r\in \{t,f\}$ and
$\Theta_R = \{\Sigma_j\Rightarrow\Pi_j\}_{1\leq j \leq m }$. An
application of $R$ is of the form:
\[\infer[]{\Gamma,\Q z(\chi[p_1],...,\chi[p_n])^{-r}\Rightarrow\Delta,\Q z(\chi[p_1],...,\chi[p_n])^{r}}
{\{\Gamma,\chi[\Sigma_j]\Rightarrow\chi[\Pi_j],\Delta\}_{1\leq j
\leq m}}\]where $\chi$ is some
$\tup{R,\Gamma\cup\Delta,z}$-mapping. Suppose that $\{\Gamma,\chi[\Sigma_j]\Rightarrow\chi[\Pi_j],\Delta\}_{1\leq j
\leq m}$ is $\M$-valid in $\tup{S,v}$. We will now show
that $\Gamma,\Q z(\chi[p_1],...,\chi[p_n])^{-r}\Rightarrow\Delta,\Q z(\chi[p_1],...,\chi[p_n])^{r}$ is
also $\M$-valid in $\tup{S,v}$. \begin{center} {\sf (a)} Let $\sigma$ be an $S$-substitution,
such that $S,v\models_\M\sigma[\Gamma]$ and
for every $\psi\in \Delta$:
$S,v{\not\models_\M}\sigma[\psi]$.\end{center}
Denote by $\widetilde{\psi}$ the $L$-formula obtained from a
formula $\psi$ by substituting every free occurrence of $w\in
Fv[\psi]-\{z\}$ for $\sigma[w]$. \\Let $\E =
\{\tup{v[\widetilde{\chi[p_1]}
\{\overline{a}/z\}],...,v[\widetilde{\chi[p_n]}
\{\overline{a}/z\}]}\ |\ a\in D\}$. We will show that $\tilde{\Q}[\E]=\{r\}$, and so
$v[\sigma[Q z(\chi[p_1],...,\chi[p_n])]]=r$. From ({\sf a}) it will follow that
$\Gamma,\Q z(\chi[p_1],...,\chi[p_n])^{-r}\Rightarrow\Delta,\Q z(\chi[p_1],...,\chi[p_n])^{r}$ is
$\M$-valid in $\tup{S,v}$. \\We prove this by showing that $\Theta_R$ is valid in some
$\E$-characteristic $\lmon$-structure. Then, by suitability of $\M$, we
shall conclude that $\tilde{\Q}_\M[\E]=r$. \\Construct the $\lmon$-structure
$\C = \tup{D',I'}$ as follows:
\begin{enumerate}[$\bullet$]
\item $D'=D$.
\item For every $a\in D$: $I'[p_i][a]=v[\widetilde{\chi[p_i]}\{\overline{a}/z\}]$.
\item For every constant $c$, $I'[c] = I[\sigma[\chi[c]]]$.

\end{enumerate}
We will now show that $\Theta_R=\{\Sigma_j\Rightarrow\Pi_j\}_{1\leq
j \leq m}$ is valid in $\C$. Suppose for contradiction that it is not so. Then there exists some $1\leq j
\leq m$, for which $\Sigma_j\Rightarrow\Pi_j$ is not valid
in $\C$. Thus
there is some $\C$-substitution $\eta$, such that:
\begin{center}({\sf b}) whenever
$p_i(\term)\in \Pi_j\cup\Sigma_j$: $p_i(\term)\in ite
(I'[p_i][I'[\eta[\term]]],\Sigma_j,\Pi_j)$.
\end{center}
We show now that $\Gamma,\chi[\Sigma_j]\Rightarrow \chi[\Pi_j],\Delta$ is not
$\M$-valid in $\tup{S,v}$, in contradiction to our assumption about the premises of the above application. \\
Let $\psi\in
ite (s,\chi[\Sigma_j],\chi[\Pi_j])$ for $s\in \{t,f\}$. Let
$\sigma'$ be the $S$-substitution similar to $\sigma$ except that
$\sigma'[\chi[y]]=\overline{a}_y$, where $a_y=I'[\eta[y]]$
for every variable $y$ occurring in $\Theta_R$. Note that $\sigma'$ is well-defined, since for every two
different variables $x,y$: $\chi[x]\neq\chi[y]$ (recall defn. \ref{candefn}). Then one of the following holds:
\begin{enumerate}[$\bullet$]
\item $\psi = \chi[p_i]\{\chi[c]/z\}$, where
$p_i(c)\in ite (s,\Sigma_j,\Pi_j)$ and $\chi[c]$ is some
term free for $z$ in $\chi[p_i]$, such that for any variable $y$ occurring in $\Theta_R$,
$\chi[y]$ does not occur in $\chi[c]$. Recall that by {\sf (b)}, $I'[p_i][I'[\eta[c]]]=s$. And so:
\[v[\sigma'[\psi]] = v[\sigma'[\chi[p_i]\{\chi[c]/z\}]]
= v[\widetilde{\chi[p_i]}\{\sigma'[\chi[c]]/z\}] =
v[\widetilde{\chi[p_i]}\{\sigma[\chi[c]]/z\}]\] 
(Recall that every variable $y$ occurring in $\Theta_R$ prevents $\chi[y]$
from occurring freely in $\Q z (\chi[p_1],...,\chi[p_n])$, and that
$\sigma,\sigma'$ only differ for variables $\chi[z]$ where $z$ occurs
in $\Theta_R$.)

By Lemma
\ref{solves}-2 and the legality of $v$:
\[v[\widetilde{\chi[p_i]}\{\sigma[\chi[c]]/z\}] =
v[\widetilde{\chi[p_i]}\{\overline{I[\sigma[\chi[c]]]}/z\}]\]
By definition of $I'$, $I'[c]=I[\sigma[\chi[c]]]$ and so:
\[v[\widetilde{\chi[p_i]}\{\overline{I[\sigma[\chi[c]]]}/z\}]=v[\widetilde{\chi[p_i]}\{\overline{I'[c]}/z\}]=
I'[p_i][I'[c]]= I'[p_i][I'[\eta[c]]]=s\]


\item $\psi = \chi[p_i]\{\chi[y]/z\}$, where
$p_i(y)\in ite (s,\Sigma_j,\Pi_j)$ and $\chi[y]$ does not occur
in $\Gamma\cup\Delta\cup \{\Q z (\psi_1,...,\psi_n)\}$ and is free
for $z$ in $\chi[p_i]$. Then $I'[p_i][I'[\eta[y]]]=s$. \\Let $a =
I'[\eta[y]]$. Then, $\sigma'[\chi[y]]=\overline{a}$ and so:
\[v[\sigma'[\psi]] = v[\sigma'[{\chi[p_i]}\{\chi[y]/z\}] =
v[\widetilde{\chi[p_i]}\{\sigma'[\chi[y]]/z\}]=\]
\[= v[\widetilde{\chi[p_i]}\{\overline{a}/z\}]=I'[p_i][a] =
I'[p_i][I'[\mu[y]]]=s\]

%
\end{enumerate}
Thus we have shown that $v[\sigma'[\psi]]=s$ whenever $\psi\in
ite (s,\chi[\Sigma_j],\chi[\Pi_j])$. Also, there is no variable $y$ occurring in $\Theta_R$,
such that $\chi[y]$ occurs in
$\Gamma\cup\Delta$, and so $\sigma[\Gamma]=\sigma'[\Gamma]$ and
$\sigma[\Delta]=\sigma'[\Delta]$. Thus for every $\psi\in \Gamma\cup
\chi[\Sigma_j]$, $v[\sigma'[\psi]]=t$ while for every $\varphi\in
\Delta\cup\chi[\Pi_j]$, $v[\sigma'[\varphi]]=f$. Hence,
$\Gamma,\chi[\Sigma_j]\Rightarrow\Delta,\chi[\Pi_j]$ is not
$\M$-valid in $\tup{S,v}$, in contradiction to
our assumption on the validity of the premises of the application above.
\\
We have shown that $\{\Sigma_j\Rightarrow\Pi_j\}_{1\leq
j \leq m}$ is valid in $\C$. Obviously\footnote{Recall that $\E =
\{\tup{v[\widetilde{\chi[p_1]}
\{\overline{a}/z\}],...,v[\widetilde{\chi[p_n]}
\{\overline{a}/z\}]}\ |\ a\in D\}$ and $I'[p_i][a] =
v[\widetilde{\chi[p_i]}
\{\overline{a}/z\}] $ for every $a\in D$ and every $1\leq i \leq n$.}, $Dist_\C = \E$. Since
$\M$ is suitable for $G$: $\tilde{\Q}_\M[\E]=\{r\}$ and so
$v[\sigma[\Q z(\chi[p_1],...,\chi[p_n])]]=r$. From this fact and
assumption {\sf (a)} it follows that $\Gamma,\Q z(\chi[p_1],...,\chi[p_n])^{-r}\Rightarrow\Delta,\Q z(\chi[p_1],...,\chi[p_n])^{r}$
is $\M$-valid in $\tup{S,v}$. \qed\medskip

\noindent {\bf Proof of Proposition \ref{mainprop}}:\\
Let $\Sss$ be a set of sequents closed under
substitution and $\Gamma\Rightarrow\Delta$ - a sequent, such that
$\Sss\cup\{\Gamma\Rightarrow\Delta\}$ satisfies the free-variable
condition. Suppose that $\Gamma\Rightarrow\Delta$ has no simple proof from
$\Sss$ in $G$. To show that
$\Sss{\not\vdash_\M}\Gamma\Rightarrow\Delta$, we will construct a structure $S$ and an $\M$-legal valuation $v$, such that
the sequents of $\Sss$ are $\M$-valid in $\tup{S,v}$, while
$\Gamma\Rightarrow\Delta$ is not. \\It is easy to see that we can limit ourselves to
the language $L^*$, which is a subset of $L$, consisting of all
the constants and predicate and function symbols, occurring in
$\Sss\cup\{\Gamma\Rightarrow\Delta\}$.
\\Let ${\bf T}$ be the set of all the
terms in $L^*$ which do not contain variables occurring bound in
$\Gamma\Rightarrow\Delta$ and $\Sss$. It is a standard matter to show that
$\Gamma,\Delta$ can be extended to two (possibly infinite) sets
$\Gamma',\Delta'$ (where $\Gamma\subseteq \Gamma'$ and
$\Delta\subseteq \Delta'$), satisfying the following properties:
\begin{enumerate}[(1)]
\item For every finite $\Gamma_1\subseteq \Gamma'$ and $\Delta_1\subseteq
\Delta'$, $\Gamma_1\Rightarrow\Delta_1$ has no
simple proof in $G$.

\item There are no $\psi\in \Gamma'$ and $\varphi\in \Delta'$, such that
$\psi\equiv_\alpha \varphi$.



\item  If $\{\Sigma_j\Rightarrow\Pi_j\}_{1\leq j \leq m}/\Q(r)$ is an $(n,0)$-ary rule of $G$
and
$\Q \  (\psi_1,...,\psi_n)\in ite(r,\Delta',\Gamma')$, then there is
some $1\leq j \leq m$, such that whenever $p_i\in ite(s,\Sigma_j,\Pi_j)
$, $\psi_i\in ite (s,\Gamma',\Delta')$ for $s\in \{t,f\}$.

\item If $\{\Sigma_j\Rightarrow\Pi_j\}_{1\leq j \leq m}/\Q(r)$ is an $(n,1)$-ary rule of $G$
and
$\Q z\  (\psi_1,...,\psi_n)\in ite(r,\Delta',\Gamma')$, then there is
some $1\leq j \leq m$, such that:
\begin{enumerate}[$\bullet$]
\item For every constant $c$,
whenever $p_i(c)\in ite(s,\Sigma_j,\Pi_j)$ for some $1\leq i \leq n$,
then $\psi_i\{\term/z\}\in ite(s,\Gamma',\Delta')$
for every term $\term\in {\bf T}$.

\item For each variable $y$, there exists some $\term_y\in {\bf
T}$, such that whenever $p_i(y)\in ite(s,\Sigma_j,\Pi_j)$ for some $1\leq i \leq n$, then
$\psi_i\{\term_y/z\}\in ite(s,\Gamma',\Delta')$.
\end{enumerate}
Note that every $\term\in {\bf T}$ is free for $z$ in $\psi_i$ for every $1\leq i \leq n$.
\item For every formula $\psi$ occurring in $\Sss$, $\psi\in
\Gamma'\cup\Delta'$.

\end{enumerate}
Note that the last condition can be satisfied because cuts on formulas
from $\Sss$ are allowed in a simple proof.

Let $S = \tup{D,I}$ be the $L^*$-structure defined as
follows:
\begin{enumerate}[$\bullet$]
\item $D = {\bf T}$.
\item $I[c] = c$ for every constant $c$ of $L^*$.
\item $I[f][\term_1,...,\term_n] =
f(\term_1,...,\term_n)$ for every $n$-ary function symbol $f$.
\item $I[p][\term_1,...,\term_n] = t$ iff
$p(\term_1,...,\term_n)\in \Gamma'$ for every $n$-ary predicate symbol $p$.
\end{enumerate}

Let $\sig$ be any $S$-substitution satisfying
$\sig[x]=\overline{x}$ for every $x\in {\bf T}$. (Note that every
$x\in{\bf T}$ is also a member of the domain and thus has an
individual constant referring to it in $L^*(D)$.)

For an $L(D)$-formula $\psi$ (an $L(D)$-term $\term$), we will denote
by $\widehat{\psi}$ ($\widehat{\term}$) the $L$-formula ($L$-term)
obtained from $\psi$ ($\term$) by replacing every individual constant
of the form $\overline{\sterm}$ for some $\sterm\in {\bf T}$ by the
term $\sterm$. More formally, $\widehat{\term}$ and $\widehat{\psi}$
are defined as follows:
\begin{enumerate}[$\bullet$]
\item $\widehat{x}=x$ for any variable $x$ of $L$.
\item  $\widehat{c}=c$ for any constant $c$ of $L$.
\item $\widehat{\overline{\term}}=\term$ for any $\term\in{\bf T}$.
\item
$\widehat{f(\term_1,...,\term_n)}=f(\widehat{\term}_1,...,\widehat{\term}_n)$.

\item
$\widehat{p(\term_1,...,\term_n)}=p(\widehat{\term}_1,...,\widehat{\term}_n)$.

\item $\widehat{\Q (\psi_1,...,\psi_n)}=\Q
(\widehat{\psi}_1,...,\widehat{\psi}_n)$.

\item $\widehat{\Q x(\psi_1,...,\psi_n)}=\Q x
(\widehat{\psi}_1,...,\widehat{\psi}_n)$.
\end{enumerate}

\begin{lem}\label{hat}Let $\term$ be an $L(D)$-term and $\psi$ -
an $L(D)$-formula.
\begin{enumerate}[\em(1)]
\item For any $z,x$:
$\widehat{\term}\{z/x\}=\widehat{\term\{z/x\}}$ and
$\widehat{\psi}\{z/x\}=\widehat{\psi\{z/x\}}$.

\item $\psi\sm \sig[\widehat{\psi}]$.

\item For every $\psi\in \Gamma'\cup\Delta'$:
$\widehat{\sig[\psi]}=\psi$.

\end{enumerate}
\end{lem}
\proof The lemma is proved by a tedious induction on $\term$ and
$\psi$.\medskip

Define the $S$-valuation $v$ as follows:
\begin{enumerate}[$\bullet$]
\item $v[p(\term_1,...,\term_n)] = I[p][I[\term_1],...,I[\term_n]]$.
\item For every $(n,0)$-ary quantifier $\Q$ of $L$,
if there is some $C\in \Gamma'\cup\Delta'$, such that
$C\equiv_\alpha \widehat{\Q (\psi_1,...,\psi_n)}$, then $v[\Q
(\psi_1,...,\psi_n)]=t$ iff $C\in \Gamma'$. Otherwise $v[\Q
(\psi_1,...,\psi_n)]=t$ iff $\tilde{\Q}[\{\tup{v[\psi_1],...,
v[\psi_n]}\}] = \{t\}$.
\item For every $(n,1)$-ary quantifier $\Q$ of $L$,
if there is some $C\in \Gamma'\cup\Delta'$, such that
$C\equiv_\alpha \widehat{\Q x(\psi_1,...,\psi_n)}$, then $v[\Q
x(\psi_1,...,\psi_n)]=t$ iff $C\in \Gamma'$. Otherwise $v[\Q
x(\psi_1,...,\psi_n)]=t$ iff $\tilde{\Q}[\{\tup{v[\psi_1\{\overline{a}/x\}],...,
v[\psi_n\{\overline{a}/x\}]}\ |\ a\in D\}] = \{t\}$.
\end{enumerate}

\begin{lem}\label{bkk}\hfill
\begin{enumerate}[\em(1)]
\item $\Is[\sig[\term]] = \term$ for every $\term\in {\bf T}$.


\item For every two $L(D)$-formulas $\psi,\psi'$: if $\psi\equiv_\alpha \psi'$, then
$\sig[\psi]\equiv_\alpha
\sig[\psi']$.

\item For every two $L(D)$-sentences $\psi,\psi'$: if
$\psi\sm\psi'$, then $\widehat{\psi}\equiv_\alpha
\widehat{\psi'}$.
\end{enumerate}
\end{lem}
\proof The claims are proven by induction
on $\term$ in the first case, and on $\psi$ and $\psi'$ in the
second and third cases.
\remove{
We show the proof for the third case:

\begin{enumerate}[$\bullet$
\item $\psi=p(\term_1,...,\term_n)$,
$\psi'=p(\sterm_1,...,\sterm_n)$, where $\term_i\sm\sterm_i$. We
will now show that $\widehat{\term}_i=\widehat{\sterm}_i$.
\\If $\term_i\sm\sterm_i$, then since $\term_i,\sterm_i$ are closed
terms, $I[\term_i]=I[\sterm_i]$. For every $1\leq i \leq n$,
one of the cases holds: 
\begin{enumerate}[$-$]
\item $\term_i,\sterm_i\in {\bf T}$. Then since
$\term_i$ is a closed term, $I[\term_i]=I[\sig[\term_i]]$. By
the first part of the lemma, $I[\sig[\term_i]]=\term_i$.
Similarly, $I[\sterm_i]=\sterm_i$ and so $\widehat{\term_i}=\term_i=\sterm_i=\widehat{\sterm_i}$.
\item $\term_i\in {\bf T}$ and $\sterm_i=\overline{\term}_i$. Then
$\widehat{\sterm_i}=\term_i=\widehat{\term_i}$.
\item $\sterm_i\in {\bf T}$ and $\term_i=\overline{\sterm}_i$.
This is symmetric to the previous case.

\end{enumerate}
Thus, $\widehat{\sterm}_i=\widehat{\term}_i$ for every $1\leq i \leq
n$. Hence, $\widehat{\psi}=
\widehat{\psi'}$, and $\widehat{\psi}\equiv_\alpha\widehat{\psi'}$.

\item $\psi = \Q (\psi_1,...,\psi_n)$, $\psi' = \Q
(\psi'_1,...,\psi'_n)$ and $\psi_i\sm \psi'_i$. By the induction
hypothesis, $\widehat{\psi_i}\equiv_\alpha\widehat{\psi'_i}$. By
Lemma \ref{alphabkk}, $\widehat{\psi}=\Q (\widehat{\psi}_1,...,\widehat{\psi}_n)\equiv_\alpha
\Q (\widehat{\psi'}_1,...,\widehat{\psi'}_n)=\widehat{\psi'} $.

\item $\psi = \Q x(\psi_1,...,\psi_n)$, $\psi' = \Q y
(\psi'_1,...,\psi'_n)$ and $\psi_i\{z/x\}\sm \psi'_i\{z/y\}$ for a fresh variable $z$.
By the induction hypothesis and Lemma \ref{hat}-1,
$\widehat{\psi_i\{z/x\}}=\widehat{\psi_i}\{z/x\}\equiv_\alpha\widehat{\psi'}_i\{z/y\}=
\widehat{\psi'_i\{z/y\}}$. By Lemma \ref{alphabkk}, $\widehat{\psi}=\Q x (\widehat{\psi}_1,...,\widehat{\psi}_n)\equiv_\alpha
\Q x (\widehat{\psi'}_1,...,\widehat{\psi'}_n) = \widehat{\psi'}$.

\end{enumerate}
}

\begin{lem}\label{gamdelt}
For every $\psi\in \Gamma'\cup\Delta'$: $v(\sig[\psi])=t$ iff $\psi\in
\Gamma'$.
\end{lem}
\proof
If $\psi = p(\term_1,...,\term_n)$, then $v[\sig[\psi]] =
I[p][I[\sig[\term_1]],...,I[\sig[\term_n]]]$. Note\footnote{This is obvious if
$\term_i$ does not occur in the set
$\{\Gamma\Rightarrow\Delta\}\cup\Sss$. If it occurs in this set, then by
the free-variable condition $\term_i$ does not contain variables bound in
this set and so $\term_i\in {\bf T}$ by definition of {\bf T}.} that for every
$1\leq i \leq n$, $\term_i\in {\bf T}$.
By Lemma \ref{bkk}-1,
$I[\sig[\term_i]]=\term_i$, and by the
definition of $I$, $v[\sig[\psi]]=t$ iff $p(\term_1,...,\term_n)\in
\Gamma'$.

Otherwise $\psi = Q (\psi_1,...,\psi_n)$ or $\psi = Q' x
(\psi_1,...,\psi_n)$. If $\psi\in \Gamma'$, then by Lemma
\ref{hat}-3 $\widehat{\sig[\psi]}=\psi\in \Gamma'$ and so
$v[\sig[\psi]]=t$. If $\psi\in \Delta'$ then by property 2 of
$\Gamma'\cup\Delta'$ it cannot be the case that there is some
$C\in\Gamma'$, such that $C\equiv_\alpha
\widehat{\sig[\psi]}=\psi$ and so $v[\sig[\psi]]=f$.\qed.

\begin{lem}
$v$ is legal in $\M_G$.
\end{lem}
\proof First we need to show that $v$ respects the $\sm$-relation. We
prove by induction on $L^*(D)$-sentences $\psi,\psi'$: if
$\psi\sm\psi'$, then $v[\psi]=v[\psi']$.
\begin{enumerate}[$\bullet$]
\item $\psi = p(\term_1,...,\term_n)$, $\psi' = p(\sterm_1,...,\sterm_n)$
and $\term_i\sm\sterm_i$ for every $1\leq i \leq n$. Then
$I[\term_i]=I[\sterm_i]$ and by definition of $v$: $v[p(\term_1,...,\term_n)] = I[p][I[\term_1],...,I[\term_n]]=
I[p][I[\sterm_1],...,I[\sterm_n]]$\\$=v[p(\sterm_1,...,\sterm_n)]$.


\item $\psi = \Q x(\psi_1,...,\psi_n)$, $\psi' = \Q y
(\psi'_1,...,\psi'_n)$ and for every $1\leq i \leq n$: $\psi_i\{z/x\}\sm
\psi'_i\{z/y\}$ for a fresh variable $z$. Then by Lemma \ref{solves}-2 for every $a\in D$:
$\psi_i\{z/x\}\{\overline{a}/z\}=\psi_i\{\overline{a}/x\}\sm
\psi'_i\{\overline{a}/y\}=\psi_i\{z/y\}\{\overline{a}/z\}$. By the induction
hypothesis,\\
$\{\tup{v[\psi_1\{\overline{a}/x\}],...,v[\psi_n\{\overline{a}/x\}]}\ |\ a\in D\}
=\{\tup{v[\psi'_1\{\overline{a}/x\}],...,v[\psi'_n\{\overline{a}/x\}]}\ |\ a\in D\}$.
One of the following cases holds:
\begin{enumerate}[$-$]
\item There is no $C\in \Gamma'\cup\Delta'$, such that $C\equiv_\alpha
\widehat{\psi}$ or $C\equiv_\alpha
\widehat{\psi'}$. Then $v[\Q x (\psi_1,...,\psi_n)]=t$
iff $\{\tup{v[\psi_1\{\overline{a}/x\}],...,v[\psi_n\{\overline{a}/x\}]}\ |\ a\in
D\}=t$ iff \\$\{\tup{v[\psi'_1\{\overline{a}/x\}],...,v[\psi'_n\{\overline{a}/x\}]}\ |\ a\in D\}=t$
iff $v[\Q y
(\psi'_1,...,\psi'_n)]=t$.

\item There is some $C\in \Gamma'\cup\Delta'$, such that $C\equiv_\alpha
\widehat{\psi}$. By Lemma \ref{bkk}-3,
$\widehat{\psi}\equiv_\alpha\widehat{\psi'}$, and so $v[\psi]=v[\psi']=t$ iff $C\in \Gamma$.

\item There is some $C\in \Gamma'\cup\Delta'$, such that $C\equiv_\alpha
\widehat{\psi'}$. Similarly to the previous case, $v[\psi]=v[\psi']=t$ iff $C\in \Gamma$.
\end{enumerate}
\item The case of $\psi = \Q (\psi_1,...,\psi_n)$, $\psi'=\Q
(\psi'_1,...,\psi'_n)$ is treated similarly.
\end{enumerate}

It remains to show that $v$
respects the interpretations of the $(n,k)$-ary quantifiers in
$\M_G$. The case of $k=0$ is not hard and is left to the reader.
We will show the proof for the case of $k=1$. Suppose by contradiction that there is some
$L^*(D)$-sentence $A=\Q z (\psi_1,...,\psi_n)$, such that
$v[A]\nin \tilde{Q}[H_{A}]$, where $H_A = \{\tup{v[\psi_1\{\overline{a}/z\}],...,v[\psi_n\{\overline{a}/z\}]}\ \ | \ a\in D\}$. From the definition of $v$, it must
be the case that\footnote{If there is no $L$-formula $C\in
\Gamma'\cup\Delta'$, such that $C\equiv_\alpha
\widehat{A}$, then by definition of $v$, $v[A]$ is always in $\tilde{\Q}[H_A]$, so this case is not possible.}:
\begin{center} ({\sf a}) there is some $L$-formula $C\in
\Gamma'\cup\Delta'$, such that $C\equiv_\alpha
\widehat{A}$, and $v[A]=t$ iff $C\in \Gamma'$. \end{center}
Suppose that $\tilde{Q}[H_A]=\{t\}$ and $v[A]=f$.
By definition of $\M_G$ and the fact that $\tilde{Q}[H_A]$ is a singleton, it must be the case that
there is some canonical rule $\{\Sigma_k\Rightarrow\Pi_k\}_{1\leq k \leq m} /\Rightarrow \Q v_1 (p_1(v_1),...,p_n(v_1))$
in $G$, such that:
\begin{center}
({\sf b}) $\{\Sigma_k\Rightarrow\Pi_k\}_{1\leq k \leq m}$ is valid in
a $H_A$-characteristic structure $\C=\tup{D_\C,I_\C}$.
\end{center}
$A = \Q z (\psi_1,...,\psi_n)$ and $C\equiv_\alpha \widehat{A}$, so $C$ is
of the form $\Q w (\varphi_1,...,\varphi_n)$. By Lemma \ref{bkk}-2, $\sig[C]\equiv_\alpha
\sig[\widehat{A}]$. By Lemma \ref{solves}-1,
$\sig[C]\sm\sig[\widehat{A}]$. By Lemma \ref{hat}-2, $\sig[\widehat{A}]\sm
A$, and thus $\sig[C]\sm A$. Let $\phi_i$ be the
formula obtained from $\varphi_i$ by substituting every $x\in Fv[\varphi_i]=\{w\}$
for $\sig[x]$. By Lemma \ref{solves}-2, $\phi_i\{\overline{a}/w\}\sm
\psi_i\{\overline{a}/z\}$ for every $a\in D$. We have already shown that $v$ respects
the $\sm$-relation, and so
$v[\phi_i\{\overline{a}/w\}]=v[\psi_i\{\overline{a}/z\}]$. Thus $H_A =
\{\tup{v[\phi_1\{\overline{a}/w\}],...,v[\phi_n\{\overline{a}/w\}]}\ |\ a\in D\}$.

Since $v[A]=f$, it follows from {\sf (a)} that $C=\Q w
(\varphi_1,...,\varphi_n)\in \Delta'$.  Then by property 3 of
$\Gamma'\cup\Delta'$, there is some $1\leq j \leq m$, such that
whenever $p_i(y)\in ite(r,\Sigma_j,\Pi_j)$, there is some $\term_y\in
{\bf T}$, such that $\varphi_i\{\term_y/w\}\in
ite(r,\Gamma',\Delta')$. By Lemma \ref{gamdelt},
$v[\sig[\varphi_i\{\term_y/w\}]]=v[\phi_i\{\sig[\term_y]/w\}]=r$.
Since $\C$ is $H_A$-characteristic, there is some $a_y\in D_\C$, such
that $I_\C[p_i][a_y]=v[\phi_i\{\sig[\term_y]/w\}]=r$.

Let us now show that $\Sigma_j\Rightarrow\Pi_j$ is not valid in
$\C$
(in contradiction to ({\sf b})). Let $\mu$ be
any $\C$-substitution, such that $\mu[y]=\overline{a}_y$ for every variable $y$ occurring in
$\Sigma_j\cup\Pi_j$. We now show that whenever $p(\term)\in
ite
(s,\Sigma_j,\Pi_j)$, $I[p][I[\mu[\term]]]=s$.\\ Let $p(\term)\in ite
(s,\Sigma_j,\Pi_j)$. If $\term$ is some variable $y$, then $I_\C[p_i][\mu[y]]=
I_\C[p_i][I_\C[\overline{a}_y]]=I_\C[p_i][{a}_y]=s$.
Otherwise $\term$ is some constant $c$. By property 3 of $\Gamma'\cup\Delta'$,
for every $\term\in {\bf T}$: $\varphi_i\{\term/x\}\in ite (s,\Sigma_j,\Pi_j)$.
By Lemma \ref{gamdelt}, $v[\sig[\varphi_i\{\term/w\}]]=v[\phi_i\{\sig[\term]/w\}]=s$.
Thus for every $\term\in {\bf T}$: $v[\phi_i\{\sig[\term]/w\}] =
v[\phi_i\{\overline{\term}/w\}]=s$. Since $\C$ is
$H_A$-characteristic, $I_\C[p_c][I_\C[c]]=s$. And so we have shown that
$\Sigma_j\Rightarrow\Pi_j$ is not valid in $\C$, in contradiction to ({\sf
b}).

The proof for the case of $\tilde{Q}[H_A]=\{f\}$ and $v[A]=t$ is symmetric.\qed

\begin{lem}For every sequent $\Sigma\Rightarrow\Pi\in \Sss$,
$\Sigma\Rightarrow\Pi$ is $\M$-valid in $\tup{S,v}$.
\end{lem}
\proof Suppose by contradiction that there is some $\Sigma\Rightarrow\Pi\in
\Sss$, which is not $\M$-valid in $\tup{S,v}$. Then there exists
some $S$-substitution $\mu$, such that for every $\psi\in \Sigma$:
$S,v\models_\M\mu[\psi]$, and for every $\varphi\in \Pi$:
$S,v{\not\models_\M}\mu[\varphi]$. Note that for every $\phi\in
\Sigma\cup\Pi$, $\widehat{\mu[\phi]}$ is a substitution instance
of $\phi$. Since $\Sss$ is closed under substitution, $\widehat{\mu[\phi]}$ also occurs in $\Sss$,
and thus by property 5 of $\Gamma'\cup\Delta'$:
$\widehat{\mu[\phi]}\in\Gamma'\cup\Delta'$. By Lemma
\ref{gamdelt},
if $\widehat{\mu[\phi]}\in \Gamma'$ then $v[\sig[\widehat{\mu[\phi]}]]=t$, and if
$\widehat{\mu[\phi]}\in \Delta'$ then $v[\sig[\widehat{\mu[\phi]}]]=f$. By
Lemma
\ref{hat}-2, $\mu[\phi]\sm \sig[\widehat{\mu[\phi]}]$. Since $v$
is $\M$-legal, it respects the $\sm$-relation and so for every $\phi\in
\Sigma\cup\Pi$: $v[\mu[\phi]]=v[\sig[\widehat{\mu[\phi]}]]$.
Thus $\widehat{\mu[\Sigma]}\subseteq \Gamma'$
and $\widehat{\mu[\Pi]}\subseteq \Delta'$. But
$\widehat{\mu[\Sigma]}\Rightarrow \widehat{\mu[\Pi]}$ has a simple proof from $\Sss$ in $G$, in contradiction to property 1 of
$\Gamma'\cup\Delta'$.

\noindent We have shown that (i) $v$ is legal in $\M$, (ii) for
every $\psi\in \Gamma'\cup\Delta'$: $v[\sig[\psi]]=t$ iff $\psi\in
\Gamma'$, and (iii) the sequents in $\Sss$ are $\M$-valid in
$\tup{S,v}$. From (ii) it follows that $\Gamma\Rightarrow\Delta$
is not $\M$-valid in $\tup{S,v}$, which completes the
proof. \qed\medskip

\noindent {\bf Proof of Proposition \ref{simple-prop}:} \hfill\break
For a set of clauses $\Theta$, denote by $\Theta\{c/x\}$ the set $\{\Gamma\{c/x\}\Rightarrow\Delta\{c/x\}\ |\ \Gamma\Rightarrow\Delta\in
\Theta\}$. Then the following lemma is easily proved:
\begin{lem}\label{concon}Let $\Theta$ be a classically consistent
set of clauses. Then for any constant $c$, $\Theta\{c/x\}$ is also
classically consistent.
\end{lem}
Now suppose that a simple canonical calculus $G$ is not
coherent. Then there is a pair of $(n,k)$-ary dual rules $R_1=\Theta_1/\Rightarrow A$ and
$R_2=\Theta_2/A\Rightarrow$, such that ${\sf Rnm}(\Theta_1\cup\Theta_2)$ is
classically consistent. If $k=0$, then the proof is similar to the
proof of theorem 4.7 in \cite{AvrLev01}.\\
Otherwise, $k=1$, $A = \Q v_1 (p_1(v_1),...,p_n(v_1))$ and
whenever $p_i(y)$ occurs in ${\sf Rnm}(\Theta_1\cup\Theta_2)$
for some variable $y$ and some $1\leq i \leq n$, there is at most
one constant $c$, such that $p_i(c)$ also occurs in
${\sf Rnm}(\Theta_1\cup\Theta_2)$. Recall that ${\sf Rnm}(\Theta_1\cup\Theta_2) = \Theta_1\cup\Theta'_2$, where
$\Theta'_2$ is obtained from $\Theta_2$ by renaming of constants
and variables which occur in $\Theta_1$ (see defn.
\ref{renaming}). We assume that the new constants in $\Theta'_2$ are in $L$ (this assumption is not necessary
but it simplifies the presentation).

Obtain the sets $\Upsilon_1,\Upsilon_2$ from $\Theta_1,\Theta'_2$
respectively as follows. For every $1\leq i \leq n$, if $p_i(c)$
occurs in $\Theta_1\cup\Theta'_2$ for some constant $c$, replace all
variables $y$, such that $p_i(y)$ occurs in $\Theta_1\cup\Theta'_2$ by
$c$ (note that this is well-defined due to the special property of
simple calculi). Otherwise, replace all variables $y$, such that
$p_i(y)$ occurs in $\Theta_1\cup\Theta'_2$ by a fresh constant $d_i$
of $L$. Then $\Upsilon = \Upsilon_1\cup\Upsilon_2$ is obtained from
$\Theta_1\cup\Theta'_2$ by replacing all variables by constants. Since
$\Theta_1\cup\Theta'_2$ is classically consistent, by repeated
application of Lemma \ref{concon}, $\Upsilon$ is also classically
consistent. Then there exists some $L$-structure $S$ in which the set
of clauses $\Upsilon$ is (classically) valid. Since $\Upsilon$
consists of closed atomic formulas, there also exists a (classical)
propositional valuation $v_S$, which satisfies $\Upsilon$. Let $\Phi =
\{A\ |\ v_{S}[A]=t,A\in \Gamma\cup\Delta, \Gamma\Rightarrow\Delta\in
\Upsilon\} $ and $\Psi = \{A\ |\ v_{S}[A]=f,A\in \Gamma\cup\Delta,
\Gamma\Rightarrow\Delta\in \Upsilon\} $.  Let $B_{j} =
\{\Pi,\Phi\Rightarrow\Sigma,\Psi\ |\ \Pi\Rightarrow\Sigma\in
\Upsilon_j \}$ for $j=1,2$. Then $B_{1}$ and $B_{2}$ are sets of
standard axioms. (Since $v_{S}$ satisfies $\Pi\Rightarrow\Sigma$,
there is some $A\in \Pi$, such that $v_{S}[A]=f$, or some $A\in
\Delta$, such that $v_{S}[A]=t$. In the former case, $A\in \Psi$ and
in the latter case, $A\in \Phi$.)

Let $x$ be a fresh variable of $L$. Define the
$\tup{R_1,\Psi\cup\Phi,x}$-mapping $\chi$ (see defn. \ref{candefn}) as
follows. For every $1\leq i \leq n$, $\chi[p_i]=p_i(x)$ if there is
some constant $c$, such that $p_i(c)$ occurs in
$\Theta_1\cup\Theta'_2$. Otherwise, $\chi[p_i]=p_i(d_i)$ (where $d_i$
is the fresh constant of $L$ chosen above). For every constant $c$ and
variable $y$ occurring in $\Theta_1\cup\Theta'_2$: $\chi[c]=c$ and
$\chi[y]=y$. It is easy to see that $\Upsilon_1 = \{\chi[\Sigma']
\Rightarrow\chi[\Pi']\ |\ \Sigma'\Rightarrow\Pi'\in \Theta_1\}$ and
$\Upsilon_2 =
\{\chi[\Sigma']\Rightarrow\chi[\Pi']\ |\ \Sigma'\Rightarrow\Pi'\in
\Theta'_2\}$. Thus the following is an application of $R_1$:
\[\infer[]{\Phi,\Q x\ (\chi[p_1],...,\chi[p_n])
\Rightarrow\Psi}
{B_{1}}\]
It is easy to check that $\chi$ is also an $\tup{R_2,\Psi\cup\Phi,x}$-mapping and so the
following is also an application of $R_2$:
\[\infer[]{\Phi
\Rightarrow\Psi,\Q x\ (\chi[p_1],...,\chi[p_n])}
{B_{2}}\]
By cut, $\Phi\Rightarrow\Psi$ is provable, but $\Phi$ and $\Psi$
are disjoint sets of atomic formulas, thus they have no cut-free
proof in $G$, in contradiction to our assumption.\qed
\vskip-50 pt

\begin{thebibliography}{Kos97}

\bibitem{Avr93}{Avron,~A.}, `Gentzen-Type Systems, Resolution and Tableaux', {\em Journal of Automated Reasoning},  vol. 10, 265--281,
1993.

\bibitem{AvrLev01}{Avron,~A.} and {I. Lev},
 `{Canonical} {Propositional} {Gentzen-type}
{Systems}', {\em  Proceedings of the 1st International Joint Conference
on Automated Reasoning (IJCAR 2001)}, R. Gore, A. Leitsch, T.
Nipkow, eds., Springer Verlag, LNAI 2083, 529-544, Springer Verlag,
2001.

\bibitem{AvrLev04}{Avron,~A.} and {I. Lev},
`Non-deterministic Multi-valued
Structures', {\em Journal of Logic and Computation}, vol. 15, 241--261,
2005.



\bibitem{BFZ94}{Baaz~M.}, {C.G. Ferm\"{u}ller} and {R.Zach},
`{Elimination of } {Cuts} in {First-order} {Finite-valued}
{Logics}', {\em  Information Processing Cybernetics}, vol. 29, no. 6,
333--355, 1994.


\bibitem{BFZ98}{Baaz M.}, {C.G. Ferm\"{u}ller}, {G. Salzer} and {R.Zach},
`{Labeled } {Calculi} and {Finite-valued}
{Logics}', {\em  Studia Logica}, vol. 61,
7--33, 1998.


\bibitem{BazL}{Baaz M.} and {A. Leitsch},
`Cut-elimination and Redundancy-elimination by Resolution',
{\em Journal of Symbolic Computation}, vol. 29(2),
149--177, 2000.

\bibitem{Carn87}{Carnielli~W.},
{`Systematization of Finite Many-valued Logics through the method of Tableaux'}, {\em Journal of
Symbolic Logic}, vol. 52 (2), 473--493, 1987.
\bibitem{Church}{Church~A.}, `A formulation of the simple theory
of types', {\em Journal of Symbolic Logic}, vol. 5, 56--68, 1940.


\bibitem{Ciabnk}{Ciabattoni~A.} and Terui K.,
`Modular cut-elimination: finding proofs or counter-examples',
{\em Proceedings of the 13-th International Conference of Logic for
Programming AI and Reasoning (LPAR06)}, LNAI 4246, 135--149, 2006.




\bibitem{Dosen}{Do\^{s}en,~K.}, `What is logic?', {\em Journal of
Philosophy}, vol. 76, 285--319, 1994.

\bibitem{End72}{Enderton,~H.},{\em `{A} {Mathematical} {Introduction} to
{Logic}'}, Academic Press, 1972.

\bibitem{Fit96}{Fitting,~M.},`{\em {First}-{Order} {Logic} and {Automated}
{Theorem}} {Proving}', 109--123, Springer, 1996.



\bibitem{Gen69}{Gentzen,~G.},  `Investigations into Logical
Deduction', in {\em The collected works of Gerhard Gentzen}
(M.E. Szabo, ed.), 68--131, North Holland, Amsterdam , 1969.

\bibitem{Han98} {H\"{a}hnle,~R.}, `{Commodious} {Axiomatization of}
{Quantifiers} in {Many-valued} {Logic}', {\em Studia Logica}, vol. 61, 101--121, 1998.

\bibitem{Hen61} {Henkin,~L.}, `Some remarks on infinitely long formulas',
{\em Infinistic Methods}, 167--183. Pergamon Press, Oxford, 1961.

\bibitem{KM64}{Kalish~D. and R. Montague},
{\em Logic, Techniques. of Formal Reasoning},
New York, Harcourt, Brace and World, Inc., 1964.

\bibitem{KM95} {Krynicki,~M.} and M.Mostowski, `Henkin Quantifiers', {\em Quantifiers: logics,
models and computation, M. Krynicki, M. Mostowski and L. Szcerba eds.}, vol. 1, 193--263, Kluwer Academic Publishers, 1995.


\bibitem{Leb01}{Leblanc~H.}, `{Alternatives}  to {Standard} {First-order} {Semantics}',
{\em {Handbook} of {Philosophical}
{Logic}}, vol. 2, 53--133, 2001.

\bibitem{Miller}{Miller,~D. and E. Pimentel}, `{Using} {Linear logic to reason about sequent systems}',
{\em Proceedings of the International Conference on Automated Reasoning with Analytic Tableaux and Related Methods (Tableaux'02)}, 2--23, 2002.


\bibitem{Most}{Mostowski,~A.}, `{Axiomatizability} of some  {Many-valued} {Predicate} {Calculi}',
{\em Fundamenta Mathematicae, }, vol. 15, 165--190, North Holland, Amsterdam , 1961.

\bibitem{Salzer96} {Salzer,~G.}, `Optimal
axiomatizations for multiple-valued operators and quantifiers based on semilattices',
{\em Proceedings of 13-th CADE, Springer, M. McRobbie and J. Slaney eds.}, vol. 1104, 688--702, 1996.

\bibitem{Shroder}{Schroeder-Heister,~P.}, `Generalized Rules for
Quantifiers and the completeness of the intuitionistic operators
$\&$, $\vee$, $\supset$, $\curlywedge$, $\forall$, $\exists$'

\bibitem{Shro}{Shroeder-Heister,~P.}, `Natural deduction calculi
with rules of higher levels', {\em
Journal of Symbolic Logic}, vol. 50, 275--276, 1985.

\bibitem{Sho67}{Shoenfield,~J.R.},{ `{Mathematical} {Logic}'},
109--123, {\em Association for {Symbolic} {Logic}}, 1967.

\bibitem{Tro}{Troelstra,~A.S.} and {H. Schwichtenberg}, {`Basic
Proof Theory'}, {\em Cambridge University Press}, 126--127, 2000.

\bibitem{Dal97}{van Dalen,~D.},{\em `{Logic} and {Structure}'},
109--123, Springer, 1997.

\bibitem{Zam_ISMVL}{Zamansky,~A. and A. Avron}, {`Quantification in non-deterministic multi-valued structures'},
{\em Proceedings of the 35th IEEE International Symposium on Multiple-Valued Logic
(ISMVL05)}, 296--301, 2005.

\bibitem{Zam_SL}{Zamansky,~A}. and {A. Avron},
`Cut Elimination and Quantification in Canonical Systems',
{\em Studia Logica (special issue on Cut Elimination)}, vol. 82(1), 157--176,
2006.

\bibitem{Zucker1}{Zucker,~J.I.}, `The adequacy problem for
classical logic', {\em Journal of Philosophical Logic},
vol. 7, 517--535, 1978.

\bibitem{Zucker2}{Zucker,~J.I. and R.S. Tragesser},
`The adequacy problem for
inferential logic', {\em Journal of Philosophical Logic},
vol. 7, 501--516, 1978.

\end{thebibliography}
\end{document}